\documentclass[times,final]{elsarticle}
\pdfoutput=1

\usepackage{jasr}
\usepackage{framed,multirow}

\usepackage{amssymb}
\usepackage{latexsym}

\usepackage{url}
\usepackage{amsmath}
\usepackage{xcolor}
\usepackage{subcaption}
\definecolor{newcolor}{rgb}{.8,.349,.1}

\usepackage[citebordercolor=white]{hyperref}

\journal{Advances in Space Research}

\begin{document}

\verso{Given-name Surname \textit{etal}}

\begin{frontmatter}

\title{Similarities and differences in accretion flow properties between GRS 1915+105 and IGR J17091-3624: a case study}
\author[1]{Anuvab \snm{Banerjee}\corref{cor1}}
\cortext[cor1]{Corresponding author: Anuvab Banerjee;
  email: anuvab.ban@gmail.com}
\author[2]{Ayan \snm{Bhattacharjee}}
\author[3]{Dipak \snm{Debnath}}
\author[3]{Sandip K. \snm{Chakrabarti}}
\address[1]{S. N. Bose National Centre for Basic Sciences, Salt Lake, Kolkata 700106, India}
\address[2]{Department of Physics, College of Natural Sciences, UNIST, Ulsan 44919, Korea}
\address[3]{Indian Center for Space Physics, 43 Chalantika, Garia St. Road, Kolkata 700084, India}


\begin{abstract}
We perform a comparative spectro-temporal analysis on the variability classes of GRS 1915+105 and IGR J17091-3624 to draw inferences regarding the underlying accretion flow mechanism. The $\nu$, as well as C2 class \textit{Rossi X-Ray Timing Explorer} observation, have been considered for analysis. We investigate the intensity variation of the source in different energy domains that correspond to different components of the accretion flow and infer the relative dominance of these flow components during the dip/flare events. We correlate the dependence of the dynamic photon index ($\Theta$)  with intensities in different energy bands and comment on the transition of the source to hard/soft phases during soft dips/flares. We also report the presence of sharp QPOs at $\sim 7.1$ Hz corresponding to both softer and harder domain in the case of $\nu$ variability class of GRS 1915+105 and discuss the possible accretion flow configuration it suggests. Sharp QPO around $\sim 20$ mHz is observed in $\nu$ and C2 classes of IGR J17091-3624 in low and mid energy band (2.0-6.0 keV and 6.0-15.0 keV), but remains undetected in high energy (15.0-60.0 keV). The 2.5-25.0 keV background-subtracted spectra have also been fitted with TCAF along with a Compton reflection component. A plausible accretion flow mechanism in order to explain the observed variability has been proposed.
\end{abstract}


\end{frontmatter}


\section{Introduction}

Black Hole X-ray Binaries (BHXBs) are the accreting systems in which a  black hole accretes hot and diffused matter from its stellar companion. A comprehensive investigation on such accreting systems provides a unique opportunity to understand the accretion process in extreme physical environments. The spectro-temporal analysis using the photons emitted from the accreting matter provides important clues regarding the underlying accretion mechanism. A majority of such objects are mostly observed in low-intensity transient states, and occasionally during outbursts, their peak luminosity increases by few orders of magnitude. The variation of spectral and timing features of black hole candidates (BHCs) during an outburst is well studied in the literature (McClintock \& Remillard 2006; Mandal \& Chakrabarti 2006, 2008, 2010; Dutta \& Chakrabarti 2010; Debnath et al. 2008, 2013; Nandi et al. 2012). An outbursting BHC can evolve through four different spectral states: the hard state (HS), hard-intermediate state (HIMS), soft-intermediate state (SIMS), and soft state (SS) (Belloni et al. 2005; McClintock \& Remillard 2006; Nandi et al. 2012; Debnath et al. 2013). Low-frequency quasi-periodic oscillations (LFQPOs) have often featured in the power density spectra (PDS) corresponding to some of these spectral states (McClintock \& Remillard 2006). All the different spectral states have also been connected to the different branches of the hardness-intensity diagrams (HIDs; Belloni et al. 2005; Remillard \& McClintock 2006; Debnath et al. 2008; Fender \& Belloni 2012) or accretion rate ratio intensity diagram (ARRID; Jana et al. 2016; Chatterjee et al. 2020).\par

Among all known Galactic BHXBs, the richest suite of structured variability patterns across the entire electromagnetic spectrum except for the visible band has been detected in the case of the black hole candidate GRS 1915+105. Since its discovery in 1992 with WATCH detector (Castro-
Tirado et al. 1992), the activity of this source has never been switched off. The estimated mass of the black hole was reported to be $14\pm{4} ~ M_\odot$ (Greiner et al. 2001). Situated at a distance $\sim 12.5$ kpc (Mirabel \& Rodr\'{i}guez 1994), the luminosity of GRS 1915+105 has often been measured to be near Eddington or super-Eddington (e.g., Done et al. 2004). Based on the HID and color-color diagram, the large number of phenomenological patterns have been divided into more than a dozen variability classes (Belloni et al. 2000; Klein-wolt et al. 2002; Hannikainen et al. 2005), such that the order of magnitude intensity fluctuation happens within a span of $\sim$seconds to $\sim$days. All these classes had been broadly categorized into three fundamental X-ray spectral states: (1) low luminosity, disk emission dominated state (soft, state A); (2) highly luminous, thermal, and/or non-thermal photon dominated state (softer, state B); and (3) low luminosity non-thermal photon dominated state (hard, state C). Rapid X-ray oscillation in highly variable classes was inferred to be the result of the rapid transition between these spectral states (Belloni et al. 2000; Yadav \& Rao 2001). The interpretation regarding the genesis of such phenomenological diversity ranged from the limit cycle oscillation of accretion-ejection events of an unstable accretion disk (Belloni et al. 1997; Mirabel et al. 1998; Neilsen et al. 2011) to high mass accretion rate around the black hole (Done et al. 2004). The variability was also connected to the oscillation between stable and unstable states driven by an input function related to the disk accretion rate (Massaro et al. 2014, 2020a,b). However, any definitive conclusion could not be obtained in the early days of investigation owing to the non-detection of another object showing similar variability characteristics. \par

In the case of another Galactic X-ray emitting black hole candidate IGR J17091–3624 discovered with \textit{INTEGRAL}/IBIS in 2003, similar rich variability patterns like GRS 1915+105 was detected (Kuulkers et al. 2003). Investigation using the archival data corresponding to different X-ray missions confirmed that IGR J17091–3624 was active in 1994, 1996, 2001, 2003, 2007 and 2011 (Revnivtsev et al. 2003; Capitanio et al. 2006, Krimm et al. 2011). The 2011 outburst of this source showed repetitive, large-amplitude intensity variations in the lightcurves similar to GRS 1915+105 (Altamirano et al. 2011a,b,c; Pahari et al. 2011). Following the criteria of defining the variability classes corresponding to GRS 1915+105 (Belloni et al. 2000), Altamirano et al. (2011c) identified $\alpha$, $\beta$, $\mu$, $\nu$, $\rho$ and $\lambda$ classes in the case of IGR J17091-3624 as well. Extended periods of non-variability typical of $\chi$ classes of GRS 1915+105 had also been detected. It had also been observed that prior to the transition to the $\rho$ class, like GRS 1915+105, IGR J17091-3624 also passes through soft/hard intermediate state (Pahari et al. 2011). In addition, some unique variability classes had also been reported in the case of IGR J17091-3624, which had never been observed in the case of GRS 1915+105 (Pahari et al. 2012). LFQPOs, as well as mHz and high-frequency QPOs (HFQPOs) had been detected during Rossi X-ray Timing Explorer (RXTE) mission (Rodriguez et al., 2011, Altamirano et al., 2011c; Altamirano \& Belloni, 2012). In particular, a 10 mHz QPO, as well as 25-30 mHz QPO in the X-ray lightcurve of RXTE/PCA observation  were identified by Altamirano et al. (2011 a,b). The discovery of a second source with similar phenomenological characteristics as that of GRS 1915+105 provides a unique opportunity to test the models used to understand the spectro-temporal behaviors of GRS 1915+105. \par 

Despite similar variability characteristics of these two sources, IGR J17091-3624 is observed to be 10-50 times fainter compared to GRS 1915+105. Assuming the variability patterns as a consequence of near Eddington limit emission, this mismatch of intensity had been attributed to either IGR J1709-3624 hosting a small black hole ($< 3~M_\odot$), or residing in more than 17-20 kpc away (Altamirano et al. 2011a). However, results from other multi-wavelength investigations are at variance with these conclusions. Contemporaneous multi-wavelength campaigns implied that the source distance lies between $11$ kpc and $17$ kpc (Rodriguez et al. 2011). Employing the photon index ($\Gamma$)-QPO frequency ($\nu$) correlation method as well as broadband spectral fitting using Two-Component Advective Flow (TCAF), the mass of  IGR J17091-3624 had been inferred to be in the range of $8.7~M_\odot-15.6~M_\odot$ with 90\% confidence level (Iyer, Nandi \& Mandal 2015). Alternative scenarios like high inclination angle of the disk plane ($i > 53^\circ$) and low black hole spin were proposed to explain the observed faintness (Capitanio et al. 2012; Rao \& Vadawale 2012). In the absence of direct measurement of the fundamental physical parameters of IGR J17091-3624, all the source parameters like mass, inclination angle and spin remain poorly constrained and the physical origin of the timing behavior of this source and its parallel with GRS 1915+105 is still debated. \par 

Prior to the launch of RXTE, the TCAF model which incorporates the high angular momentum Keplerian flow and low angular momentum sub-Keplerian flow in the same framework (Chakrabarti 1995, 1997) was proposed for both Galactic and extragalactic black holes. This model has been successfully applied to explain the observed spectral features of transient X-ray sources and infer the underlying accretion flow dynamics (Debnath et al. 2014, 2015a,b, 2017, 2020; Mondal et al. 2014, 2016; Jana et al. 2016, 2017, 2020; Molla et al. 2016, 2017; Chatterjee et al. 2016, 2019, 2020; Bhattacharjee et al. 2017). Recently, the TCAF paradigm has been successfully applied to model the spectro-temporal characteristics of weakly magnetized neutron stars as well after incorporating the modifications corresponding to the inner boundary (Bhattacharjee \& Chakrabarti, 2017, 2018, 2019, 2020, 2021). In TCAF paradigm, the `Compton cloud', namely the repository of hot electrons is formed naturally behind the centrifugal barrier because of the pile-up of the high angular momentum Keplerian disk matter and conversion of kinetic energy to thermal energy beyond the centrifugal barrier. This hot, puffed-up region which intercepts the soft seed photons from the Keplerian disk and reprocesses them via inverse Comptonization is known as the CENtrifugal pressure supported BOundary Layer (CENBOL). Thus, in the TCAF paradigm, the low energy blackbody hump, as well as the high energy power-law tail, are incorporated in a single framework. It had also been demonstrated that under suitable flow conditions, outflows can be ejected from the CENBOL region transverse to the disk plane and the rate of ejection depends on the shock compression ratio (Chakrabarti 1999). The capability of the CENBOL region to control outflow rates had been investigated using smoothed particle hydrodynamics simulation technique by Molteni, Lanzafame \& Chakrabarti (1994). The theoretically predicted variation of outflow rate with the shock strength had later been confirmed using extensive hydrodynamic simulations as well (Garain et al. 2012). Very recently, the interplay between accretion and ejection under the TCAF paradigm has been successfully applied to understand the spectral and timing features of the non-variable $\chi$ class (Banerjee et al. 2020a) and the highly variable $\theta$ class (Banerjee et al. 2020b) of GRS 1915+105. \par 

It had been demonstrated that the variability classes of GRS 1915+105 can be well characterized by invoking the Comptonizing Efficiency (CE) parameter, which is the ratio of the power-law photons and intercepted soft seed photons averaged over the duration of a variability class (Pal et al. 2013). These CE values carry information regarding the geometry of the accretion flow configuration. A large CE value would imply a large Compton cloud, which is observed in the hard state. Pal et al. (2015) subsequently had demonstrated that corresponding to the visually `similar looking' variability classes of GRS 1915+105 and IGR J17091-3624 both, the CE values turn out to be roughly similar, despite the fact that the mass of these two objects can be totally different. All the variability classes of these two objects could also be arranged in the same way in ascending order of CE values (Pal et al. 2015).  Therefore, the description of variability in terms of CE values turns out to be a mass-independent description and the likely behavior of these two sources in terms of CE implies that the accretion flow configuration evolves in the same way with the time evolution of sources. This further establishes the picture of an accretion flow configuration that encompasses the disk component and the Compton cloud together as essential components of accretion flow as envisaged in TCAF, instead of treating them as separate entities. \par 

In this paper, we attempt to provide a plausible accretion flow picture that underlies the observed spectral and temporal variation in two of the variability classes corresponding to these two sources. We show that once the presence of two accretion flow components is granted, there is a way of explaining the variability properties as a consequence of the interplay between the flow components. This can be taken to be the first step of a larger program of undertaking extensive wide band spectro-temporal analysis and making comparative studies for all the major variability classes detected in these two sources and understanding the holistic accretion flow picture that such a study reveals.

\section{Data Selection and Analysis Procedure}
Since the beginning of the 2011 outburst in February, IGR J17091–3624 had been observed with the Proportional Counter Array (PCA; Jahoda et al. 2006) onboard RXTE almost on a daily basis. For the purpose of our comparative study on the basis of detailed spectro-temporal analysis, we have chosen RXTE pointed observations corresponding to GRS 1915+105 and IGR J17091–3624 both, which show significant and rapid variability. The variability patterns corresponding to some of these observations of IGR J17091–3624 had been pointed to be similar to those observed in the case of GRS 1915+105. \par 
 
 The IGR J17091–3624 observation 96420-01-05-00 (MJD 55648) and GRS 1915+105 observation 10408-01-40-00 (MJD 50369) pertain to the $\nu$ class, as identified by Altamirano et al. (2011d) on the basis of the classification of Belloni et al. (2000). The $\nu$ class is characterized by the quasi-periodic appearance of slow variability over few tens of seconds, followed by short duration flares and subsequent sharp dips. In the case of IGR J17091-3624, the dip-to-dip duration is $\sim50$ second and the intensity gradually rises from dip to flare. In the case of GRS 1915+105, on the other hand, the flares are sharper and single-peaked with the dip-to-dip duration of $\sim$100 second As the intensity gradually recovers from the dip, the photon count remains nearly constant for few tens of seconds, and then prior to the flare the photon count sharply increases. \par 
 
 On the other hand, IGR J17091-3624 observation 96420-01-31-05 (MJD 55832) shows the recurrent appearance of nearly non-variable and highly variable sub-states and such variability is hitherto unobserved in the case of GRS 1915+105 (Pahari et al. 2012). This variability class is known to be the C2 class. Following the persistence of the variable sub-state for a few hundred seconds, the source makes a transition into the non-variable sub-state in a few tens of seconds. The mean intensities of the two sub-states are observed to be similar.\par 
 
For the purpose of our case study, we have chosen these publicly available RXTE pointed observations to undertake the spectral and temporal analysis. Standard data reduction and background estimation procedures as described in Debnath et al. (2013, 2015a) have been followed using the XRONOS software package HEASOFT version 6.25. In order to perform the spectral analysis, standard-2 mode data from the PCA had been considered. The 2.5-25.0 keV background-subtracted spectra from all layers of individual Proportional Counter Units (PCUs) from all the available PCUs are co-added to enhance the photon count rate. The spectral model fitting of PCA spectra has been accomplished using HEASOFT's spectral analysis software XSPEC version 12.10. In order to account for the interstellar absorption, photoelectric absorption model \textsc{phabs} has been used and the hydrogen column density corresponding to GRS 1915+105 and IGR J17091-3624 are kept fixed at $6.0 \times 10^{22}~ \text{cm}^{-2}$ (Muno et al. 1999) and $1.1 \times 10^{22}~ \text{cm}^{-2}$ (Krimm \& Kennea 2011) respectively.  The 1\% systematic error has been employed to achieve the best-fit parameters. Mass of the black hole is not a flow parameter, rather an intrinsic property of the black hole. Therefore, for the purpose of our spectral fitting with TCAF, the mass of the black hole has been kept fixed. Mass of GRS 1915+105 is pegged at $14 ~ M_\odot$, and mass of IGR J17091-3624 is kept frozen at $10 ~ M_\odot$.\par 

 In order to accomplish the timing analysis, RXTE/PCA Good Xenon mode data with $\sim$ 0.953 microsecond time resolution has been used. In case Good Xenon file containing the data from all the channels is not available, we have combined the time series obtained from event mode and binned mode data. The binned mode data available for 0-35 channels are extracted using FTOOLS task SAEXTRCT, and are combined with the event mode data for the rest of the channels after extraction using SEEXTRCT task. The addition of the lightcurves is performed using `lcmath' task. The Power Density Spectra (PDS) are generated using FTOOLS task \textsc{powspec}. The normalization parameter has been chosen to be -2 to subtract the expected white noise background to provide the squared \textit{rms} variability per unit frequency interval. The lightcurves are re-binned to 0.01 second in order to obtain the Nyquist frequency of 50 Hz. FTOOLS task \textsc{timetrans} is employed to select a time segment of interest to produce the lightcurves and corresponding PDS over that specific time interval. The PDSs are rebinned by a geometric rebinning factor of 1.03 in the frequency space in order to improve the signal-to-noise ratio. A detector deadtime correction was not applied on our data, since we found the typical correction factor to be quite moderate ($\sim$ 1.05), which leads to only a marginal change in exposure and normalization factor). \par 
 
 In the case of the rapidly varying classes of our interest, sometimes it becomes difficult to perform the spectral fit or obtain statistically significant PDS during the individual dip and flaring events. In such cases, we follow the intensity ratio method used in Vadawale et al. (2001a, 2003a). The full intensity of the source is obtained from 2.0-60.0 keV background-subtracted lightcurve produced from raw science files. In order to investigate the presence/absence of the soft/hard components during individual dip/flare events, the lightcurve has been further sub-divided into the soft, intermediate and hard energy intervals defined as 2.0-6.0 keV (band A, 0-14 channels), 6.0-15.0 keV (band B, 15-35 channels) and 15.0-60.0 keV (band C, 36-100 channels). The source characteristics during the dips/flares are monitored in terms of the ratios of individual bands with broadband intensity (2.0-60.0 keV). In case the quasi-stable intensity variation over extended time intervals (few tens of seconds) is observed in either of the sources, the appearance/disappearance of QPOs in different energy bands are investigated and QPO properties are measured. The dynamic PDS are also produced to monitor the time evolution of peaked components in PDS with the intensity variations. \par 
 
 We have also used the methodology followed by Ghosh et al. (2018) employing dynamic photon index ($\Theta$) to understand the source behavior during transition regions. If the photon counts in A, B $\&$ C energy bands are denoted as a, b and c respectively, then $\Theta$ is defined as

 \begin{equation}
     \Theta = \tan^{-1}[\frac{c-b}{\mid E_C-E_B \mid}],
 \end{equation}
where $E_B (=10.5 \text{keV})$ \& $E_C (=37.5 \text{keV})$ are the mean energies corresponding to B and C blocks. Therefore, $\Theta$ effectively denotes the spectral slope of the hard energy domain drawn on a linear scale. The mathematical domain of excursion of $\Theta$ is $-1.57 < \Theta < +1.57$, even though $\tan{\Theta}$ would assume mostly the negative values since in most of the cases $c$ is expected to be lesser compared to $b$. Therefore, in the softest states $\Theta \rightarrow -1.57$, while in harder states $\Theta \rightarrow 0$. Since the $C$ block covers a wider energy interval compared to $B$, it is also possible to obtain positive $\tan{\Theta}$ values in harder states. Therefore, the temporal evolution of $\Theta$ becomes instrumental in inferring the spectral state evolution with the intensity variation of the source (Ghosh \& Chakrabarti 2018, Ghosh et al. 2019). This index has been constructed in such a way that it is most responsive around the state transition. Since $\tan^{-1}$ maintains a sigmoid profile, this index is most responsive around the inflection point where C-band intensity crosses B-band intensity (around $c \ge b$, i.e. harder domain). In the low luminosity sources with rapid variability (like IGR J17091-3624) where our ability to monitor the spectral state evolution using the conventional method of power-law fit is seriously limited by the low S/N ratio of the time-resolved spectra, dynamic photon index can serve as a credible indicator of source state evolution. We utilize this feature in our analysis to trace the softness/hardness of the corresponding source as it evolves through successive spikes/dips.

 \section{Results}
 We describe below the results obtained from the spectral and temporal analysis on all the different variability classes. Since all the classes differ in their morphologies and intensity patterns, the strategy of the analysis differ from one class to another as well.
 
 \subsection{The $\nu$ Class}

Fig. 1(a,b) features the $\nu$ class lightcurves corresponding to GRS 1915+105 and IGR J17091-3624 respectively. In the case of GRS 1915+105, the photon count remains almost steady for $\sim$ 70-80 second, and then short duration ($\sim 10$ second) flare is observed where the photon count increases by a factor $\sim$ 2 (Fig. 1a). Subsequently, a sharp dip is followed where the intensity decreases by a factor $\sim$ 3. Subsequent to the dip, the photon count recovers within a few tens of seconds when the second flare of magnitude comparable to the previous one is detected, after which the steady intensity phase is followed. In the case of IGR J17091-3624, however, the flaring events are much more repetitive ($\sim$ 40-50 second) and in between the flares, the intensity steadily increases. The post-flare intensity dip is also observed in the case of IGR J17091-3624, although the post-dip recovery of photon intensity is quicker in this case. The steady photon intensity phase is almost unobserved in IGR J17091-3624. \par 
 
In order to investigate the energy dependence of the source evolution, we plot the photon counts of the two sources in 2.0-60.0 keV as well as A, B and C bands in Fig. 2(a,b). In the case of GRS 1915+105, the intensity in A and B bands evolve almost in the same way as the total intensity in 2.0-60.0 keV evolves (Fig. 2a, panel a, b and c). However, in the C band, the sharp spikes during flaring events are absent and sharp reductions of photon intensity are observed during the overall intensity dips.
This confirms the evolution of the intensity across all three energy bands. However, in the case of IGR J17091-3624, the intensity in the C band does not show any rising or declining pattern, although distinguishable intensity variation is detected in the other two bands (Fig. 2b). \par 

In order to obtain the relative changes of photons in different energy bands, we plot the ratio of photon counts A, B and C bands with the total number of photons (Cnt) in full 2.0-60.0 keV energy range (Fig. 3a-b). From Fig 3(a), it is observed that in the case of GRS 1915+105, the ratio of A band intensity and cnt ($r\_A = A/\text{cnt}$) gradually increases as the source evolves from dip to flare, but the ratio of B band intensity and cnt ($r\_B = B/\text{cnt}$) increases during the dips. The inverted shapes of $r\_B$ and $r\_C$, on the other hand, imply that the intensity in C-band preferentially decreases during the dips and does not increase significantly during the flares as well. However, the absolute values of photon intensities in these three energy bands are significantly different. From Fig. 2(a), we observe the intensity in the C band is almost one order of magnitude smaller compared to the A or B band in the case of GRS 1915+105. Therefore, in order to further consolidate our observation, the normalized counts in all the energy bands are plotted and their evolution has been observed as a function of time (Fig. 4(a,b)). From Fig. 4(a), it is further observed that the C band intensity does not change during the flares, but gets preferentially decreased during the dip intervals. Although the intensity variation during the flares is the most prominent in the soft energy domain (A band), the B band intensity also changes by $\sim1.5$ during these intervals. The preferential decrease of the C band during intensity dips suggests the plausible disappearance of a hard tail during those dips.  \par 

In the case of IGR J17091-3624, on the other hand, both the A and B band intensities change significantly (by a factor $\sim2-3$), but the C band intensity fluctuates around a mean value $\sim 40 ~ \text{counts}~ \text{s}^{-1}$ (Fig. 2(b)). In Fig 3(b), we observe the occurrence of peaks in $r\_A$ during the same instants at which broadband peaks are observed. This indicates the dominance of disk flux during the individual micro-flares. The $r\_B$ does not show sharp peaks, suggesting that the preferential increase of B-band intensity is potentially lesser compared to A-band. The dips in the C-band, on the other hand, are merely a consequence of nearly steady intensity across the temporal evolution of the source. Fig. 4(b) shows that the A and B band normalized counts evolve almost the same way from dip to flare, but the C band normalized counts do not show recurrent dips and flares.  \par 

In order to further check the softness/hardness of the respective sources during their time evolution, the dynamic photon index ($\Theta$) has been plotted along with the total intensity variation (2.0-60.0 keV) of the sources (Fig. 5(a,b)). In the case of GRS 1915+105, the photon count in the C-band is smaller than B-band by almost one order of magnitude (Fig. 4a), and this leads to $\Theta < -1.5$. In this case, $\Theta$ does not make large excursions, but the relative hardness of the source during the intensity dips and gradual softness towards the flaring events is quite conspicuous (Fig. 5(a)). In the case of IGR J17091-3624, however, $\Theta$ makes sharp jumps from $\sim0$ to $\sim -0.5$ during the flares. In this case, the proportion of the C-band count is substantially larger compared to GRS 1915+105. Therefore, while the variation of the hard component can play a crucial role in the $\nu$ class evolution of GRS 1915+105, in the case of IGR 17091-3624, this harder component is stable and dominant. This observation is further strengthened by the intensity vs. $\Theta$ plots (Fig. 6(a,b)). In Fig. 6(a), panels b and c, we observe that in the case of GRS 1915+105, the increase in photon intensity in A and B bands both lead to marginally lower $\Theta$ values, namely increase in `softness'. In panel d, such a relationship is not so apparent, but we observe clustering of the majority of observations within a narrow range of $\Theta$. However, in the case of IGR 17091-3624, the photon count in C-band and $\Theta$ values are positively correlated, as depicted in Fig. 6(b), panel (d). The best-fitted line in panels (b,c) suggests softening of the source with the intensity of counts in A and B bands both, but in panel (d), the best-fitted line suggests that the source `hardens' as count increases in band C.\par 

In order to investigate the presence of QPOs during the different phases of the source evolution, we have produced the dynamic PDS corresponding to the time series in the entire 2.0-60.0 keV domain, as well as the A, B and C bands. The dynamic PDS has been generated by producing individual PDS corresponding to the time series of 50-second intervals and making the origin of each successive interval advance by 1 second. Corresponding to a 150-second long segment containing flaring events and successive relaxation to steady intensity phase, we provide the dynamic PDS in Fig. 7(a-d). During the initial flaring phase, we observe no signature of the peaked component in the dynamic PDS, but during the steady phase, a prominent peaked component around $\sim 7$ Hz is observed. This peaked component is present in all A, B and C energy bands around the same frequency, suggesting the oscillation of the Comptonizing component and additional power-law component simultaneously. We have examined the individual PDS corresponding to the time series in full 2.0-60.0 energy range as well as all the energy bands A, B and C. Sharp QPOs are observed in all the cases. In order to quantify the QPOs, the PDS are fitted using two Lorentzians and one power-law profile. Corresponding to the full 2.0-60.0 keV lightcurve, the centroid frequency of the fundamental QPO has been found to be $7.19 \pm 0.03$ Hz, with the fractional rms power to be $4.55 \pm 0.13\%$ (Fig. 8(a)). We observe that in the case of band-A, the fractional rms power corresponding to the fundamental QPO is $\sim 3.2\%$, but in band B and C the rms amplitude increases to $\sim$ 5.2\% and $\sim$ 9.0\% respectively. Such energy dependence of rms power contained in QPO has been reported in the case of GRS 1915+105 earlier as well (Chakrabarti \& Manickam 2000). The QPO normalization also increases from softer domain (A band) to harder domain (C band). Table 1 contains a full list of fitted parameters. \par

In the case of IGR J17091-3624, on the other hand, we observe very sharp QPO in the PDS around $23$ mHz. We fit the fundamental QPO using the Lorentzian profile. The fundamental QPO frequency turns out to be $23.08\pm 0.04$ mHz, corresponding to which the fractional rms power was $\sim14\%$ (Fig. 9(a)). Similar to GRS 1915+105, we attempt to observe the signature of this prominent mHz QPO in different energy segments. Prominent mHz QPO in the same frequency is observed in band A and B (Fig. 9(b,c)), but in band C the QPO appears to be less prominent and the broadband noise component becomes dominant (Fig. 9(d)). This energy dependence of QPO is at variance with the observed variation in GRS 1915+105, where sharp QPOs in all A, B and C bands are detected. Lower QPO frequency in IGR J17091-3624 suggests a higher length scale of oscillation, which in the context of accretion flow happens to be the inner edge of the disk and the boundary of the Comptonizing region. 

\par 

In order to understand the similarities/differences of these two objects in terms of their spectral characteristics, we first attempted the fitting of the spectra using empirical models like multicolor disc blackbody component (MCD, model as \textsc{diskbb} in XSPEC; Mitsuda et al. 1984) and \textsc{power-law}. In the case of GRS 1915+105, the \textsc{diskbb+power-law} fitted does not yield acceptable spectral fit for $\sim$ 200 second long quasi-steady intensity intervals between dip and flare events. The $\chi^2/\text{DOF}$ turns out to be 156.70/49. The profile was unable to account for the higher energy tail of the spectra ($> 20$ keV) and excess emission is observed $\sim 6.5$ keV. The addition of a Gaussian line does not bring reduced $\chi^2$ below 2.0. The replacement of the power-law model with the Comptonization model \textsc{nthcomp} leads to significant improvement of the quality of spectral fit ($\chi^2/\text{DOF}$ = 48.35/45). During this spectral fitting, we tie the disk blackbody temperature ($T_{\text{in}}$ = $1.20 \pm 0.03$ keV) of \textsc{diskbb} model with the input soft seed photon temperature of \textsc{nthcomp}. The electron temperature of \textsc{nthcomp} model turns out to be $24.3 \pm 4.0$ keV. We have calculated the energy flux contributed by the disk and the Comptonizing component separately using the \textsc{xspec} convolution model \textsc{cflux}. The \textsc{diskbb} flux ($4.89\times10^{-8} ~ \text{ergs s}^{-1}\text{cm}^{-2}$) turns out to be greater compared to the Comptonized flux contributed by \textsc{nthcomp} ($2.69\times10^{-8} ~ \text{ergs s}^{-1}\text{cm}^{-2}$). However, the energy flux of IGR J17091-3624 is one order of magnitude smaller than GRS 1915+105.
\par 
 
In case of IGR 17091-3624 the spectra were well fitted with $\textsc{diskbb+po}$ along with a gaussian at 6.5 keV ($\chi^2/\text{DOF} = 47.92/43$) to account for the Iron line emission. The energy flux pertaining to \textsc{diskbb} ($2.43\times10^{-9} ~ \text{ergs s}^{-1}\text{cm}^{-2}$) and \textsc{power-law} ($2.48\times10^{-9} ~ \text{ergs s}^{-1}\text{cm}^{-2}$) are separately calculated and are found to be almost equal. This implies that the proportion of high energy flux in case of IGR J17091-3624 is greater compared to GRS 1915+105 and this makes IGR J17091-3624 spectrally harder. For a further consistency check, we plot the $\Theta$-intensity correlation (Fig. 6(a,b)) as well, where it is apparent that in the case of GRS 1915+105, $\Theta$ is negatively correlated with the intensity in C-band (hard band), but in the case of IGR J17091-3624, it is positively correlated. Thus, the higher proportion of C-band photons primariliy governs the `hardness' of the spectra in the case of IGR J17091-3624, and this is in agreement with the results obtained from \textsc{diskbb+po} spectral fit. \par 

In order to derive further insights regarding the accretion flow geometry and investigate the presence of different spectral components, we have attempted spectral fitting using TCAF and extract the underlying accretion flow parameters. The spectral fitting using TCAF as the only model provided unacceptable fits ($\chi^2/\text{DOF}$ = 419.36/45). Both in low and high energy domain the spectra were not well fitted. The addition of a Gaussian at 6.5 keV did not provide statistically viable fits either. This leads us to investigate the presence of the Compton reflection component to account for the fluorescence line emission and its Comptonization. The spectral model employed for this purpose is the reflection of the exponentially cut-off power-law spectrum from neutral disk medium (\textsc{pexrav} model, Magdziarz \& Zdziarski 1995). The Iron abundance had been taken to be of solar value, considering the presence of the source in the Galactic bulge. The spectral fit was found to be statistically viable ($\chi^2/\text{DOF}$ = 45.13/41). Among the TCAF fitted parameters, the sub-Keplerain halo rate ($0.25~\dot{\text{M}}_\text{Edd}$) was found to be more compared to the disk rate ($0.12~\dot{\text{M}}_\text{Edd}$), which is in agreement with the `harder' nature of the source in this class. Table 2 features the full list of spectral fitted parameters. 
\par 

In case of GRS 1915+105, however, \textsc{TCAF+pexrav} model was not sufficient to bring $\chi_{\text{red}}^2$ below 2. A few of the parameters could not be constrained and the fit was unacceptable ($\chi^2/\text{DOF}$ = 764.29/44). In order to obtain a reliable fit, we added an additional power-law component arising out of secondary Comptonization/synchrotron emission from micro ejections. Since the standard power-law component does not reproduce the high energy tail of the spectra reliably, we added the \textsc{cutoffpl} component which contains the power-law photon index ($\alpha$) and an exponential roll-over factor ($\beta$). Low $\alpha$ and high $\beta$ implies relative hardening of the spectra. With this additive \textsc{cutoffpl} model, very good reduced $\chi^2$ value could be achieved ($\chi^2/\text{DOF}$ = 30.44/42). In order to make a comparison with the spectral parameters arising out of the spectral fitting of IGR J17091-3624, we froze the TCAF normalization value at the one obtained from IGR J17091-3624 fitting. The disk accretion rate $1.02~\dot{\text{M}}_\text{Edd}$ and sub-Keplerian halo rate $0.30~\dot{\text{M}}_\text{Edd}$ both turned out to be significantly greater compared to IGR J17091-3624. These relative changes are in tune with the observed luminosity differences between these two sources. In the case of GRS 1915+105, the relative accretion rates further confirm the softness of GRS 1915+105 during $\nu$ class, during which the shock location moves considerably inwards; a result that has earlier been observed for many black hole transients (see Bhattacharjee et al. 2017 for an example) as well as for $\chi$ and $\theta$ classes of GRS 1915+105 (Banerjee et al. 2020a,b). The reflection fraction is smaller in GRS 1915+105, as the spectral fits suggest. In Table 2, the spectral fitted parameters along with 90\% confidence ranges corresponding to each parameter have been provided. Fig. 10(a,b) features the 2.5-25.0 keV RXTE PCA unfolded spectra of GRS 1915+105 and IGR J17091-3624 respectively as well as the residuals after spectral fitting. 

 \subsection{The C2 class}
Having investigated the spectro-temporal features in $\nu$ variability class of these two sources, we were interested to explore the plausible signatures of these findings in the C2 class of IGR J17091-3624, which has no morphological counterpart corresponding to GRS 1915+105 (Pahari et al. 2012). This class is distinguished by alternating emergence of quasi-variable sub-state and rapidly variable sub-state, each of which is of $\sim$ a few hundred seconds duration (Fig. 11). The transition from the highly variable state to the nearly non-variable sub-state happens within $\sim$ tens of seconds. Even though within one variable sub-state the intensity excursions can be by a factor $\sim$ 4-5, the average intensities of these two sub-states remain almost the same, such that they have similar mean Hard color (Pahari et al. 2012). In order to explore the contribution of the photons from different energy bands during the small dips and flares during the variable sub-state as well as the dominance of different energy bands during the sub-state transition, we produced the energy-resolved lightcurves as done previously for $\nu$ variability class (Fig 12). The lightcurves pertaining to A and B bands show correlated behavior in terms of the appearance of dip and flare events observed in the lightcurve over the full PCA energy range (2.0-60.0 keV). However, the C-band lightcurve does not show any prominent variability and the count remains quasi-steady during the entire passage of the highly variable sub-state. The relative dominance of these three bands is further clarified in Fig. 13, where the temporal variation of the normalized counts in all three bands A, B and C are plotted. The preferential decrease in B-band count during the dip events as observed (Fig. 13) implies the plausible evacuation of the Comptonizing region during these events, i.e. ejection of a portion of the Comption cloud. This can be triggered by various physical effects like enhancement of the soft seed photons as a result of heightened disk accretion rate and consequent cooling of the outflow base leading to the blobby jet ejections. This can also be triggered by magnetic process where a portion of the Compton cloud is ejected as a result of magnetic tension effect. This scenario was invoked earlier as well by Vadawale et al. (2003a) in order to explain the radio emission from $\beta$ and $\theta$ classes of GRS 1915+105. However, the C-band count remains almost non-variable and the mean normalized intensity of C-band appears to be more as compared to the mean normalized intensity of A and B. This observation is indicative of the fact that there is almost no signature of the preferential changes of C-band intensity during this class. \par
 
In order to further test the changes in Hardness of the source during the variable to non-variable transition, we plot the dynamic photon index as a function of time (Fig 14a,b). During the dips, the $\theta$ value becomes $> 0$, which was not observed in the case of $\nu$ class (Fig. 14(a)). This is in alignment with the finding that the C band intensity remains almost unaltered during the dips while B-band experiences a preferential intensity decline. However, we observe that as the source makes a transition from variable to non-variable sub-state, there is no sharp change in $\Theta$ value (Fig. 14(b), panel (b)). This observation is in consonance with the conclusion by Pahari et al. (2012) that the sharp Hard color change is unobserved during the sub-state transition. The energy-resolved intensity dependence with $\Theta$ is shown in Fig. 15, where the variation of $\Theta$ with the full broadband intensity (2.0-60.0 keV) with the A, B and C band intensities are plotted and linear best-fit lines are drawn. We observe negative correlations of $\Theta$ with photon intensity in A and B bands, but a weakly positive correlation in the C band. This further reinforces the previous observation of the dominant normalized count in C bands across dips and flares. \par

 As in the case of $\nu$ class, we have investigated the presence of low frequency QPOs in PDS as well. The QPO observed in this case does not appear to be as sharp as observed in $\nu$ class. The Lorentzian fit of the peaked component yields the frequency of the fundamental QPO to be $22.34$ mHz, and the fractional rms power corresponding to the QPO happens to be $\sim 6.5\%$ (Fig. 16(a)). Prominent QPOs are observed in A and B band (2.0-6.0 keV and 6.0-15.0 keV respectively) as well (Fig. 16(b,c)), but in C band we do not observed QPOs (Fig. 16(d)). This low frequency QPO again suggests the larger length scale of oscillation. \par 
 
 In order to connect these temporal features with spectral characteristics and compare them with the spectral results from $\nu$ class analysis, we have undertaken the spectral analysis of steady and variable sub-states separately. In both of the cases, the combination of \textsc{diskbb} and \textsc{power-law} models along with the Gaussian emission line at 6.5 keV produced acceptable spectral fits ($\chi_{\text{red}}^2 \sim 1$). The blackbody and power-law fluxes are found to be comparable ($\sim 1.4 \times 10^{-9}~ \text{erg cm}^{-2}\text{s}^{-1}$) in both of the sub-states. In order to derive further insights regarding the accretion flow configuration, we attempted to fit the spectra using TCAF as well, with additional models to achieve the statistically significant spectral fits. In the non-variable state the spectral fit solely using TCAF rendered unacceptable results ($\chi^2/\text{DOF} = 535.20/45$). However, with the addition of the reflection model \textsc{pexrav} as before, viable spectral fit could be achieved ($\chi^2/\text{DOF} = 40.99/41$). The disk accretion rate ($0.57~\dot{M}_{\text{Edd}}$) was found to be considerably greater compared to the halo rate ($0.25~\dot{M}_{\text{Edd}}$), which makes the shock location move close to the black hole ($15 r_S$). In the case of variable phase, the spectral fit could again be achieved using $\textsc{TCAF+pexrav}$ combination ($\chi^2/\text{DOF} = 58.91/41$). The disk rate again dominates over the halo rate ($0.54\dot{M}_{\text{Edd}}$ and $0.23\dot{M}_{\text{Edd}}$ respectively) and shock location moves slightly outwards (27.15 $r_S$) compared to the steady case. Table 3 contains a full list of spectral fitted parameters. In Fig. 17(a,b), we plot the 2.5-25.0 keV RXTE PCA unfolded spectra of non-variable and highly variable phase of C2 class respectively along with the residuals after spectral fitting.

 \section{Discussions and Concluding Remarks}
 In this paper, we have attempted to undertake a case study to compare the variability properties of GRS 1915+105 and IGR J17091-3624. For this purpose, spectro-temporal analysis has been performed on $\nu$ class data of these two sources, as well as the C2 class of IGR J17091-3624 which contains both variable and non-variable phases and which has no counterpart in GRS 1915+105 morphology. From the analysis, we can draw the following conjectures regarding the underlying accretion flow scenario corresponding to these two classes.
 
  \begin{enumerate}
     \item The $\nu$ class data of IGR J17091-3624 shows quasi-periodic occurrences of intensity dips and flares. The rising profile shows an exponential rise and is followed by a fast decay. However, a closer examination of the time evolution of $\nu$ variability class across different energy bands (A, B and C) shows that the C-band intensity remains nearly unchanged, while the A and B bands show dips and flares in tune with the broadband lightcurve (2.0-60.0 keV) of the source (Fig. 2(b)). Since the accretion flow components contributing to the A and B band photons are mostly variable, this could be attributed to the rapid evacuation and successive replenishment of the disk matter as well as a portion of the Comptonizing region. Such dynamic removal and refilling of the inner disk matter could be attributed to, as envisaged by Capitanio et al. (2012), the thermal-viscous instability driven by the radiation pressure in the hot inner disk region (Lightman \& Eardley 1974; Nayakshin \& Rappaport 2000). It could also be connected to the so-called `magnetic rubber-band effect', where the blobby evacuations of the hot inner disk material are attributed to the catastrophic collapse of the stochastic magnetic flux tubes (Nandi et al. 2001). Using hydrodynamic simulations, it has been demonstrated that the radiation pressure and heating driven inner disk instability can also launch variable winds of different amplitude and time-scale (Janiuk et al. 2015, A\&A, 574, 92). It is, however, outside the purview of the present manuscript to explore the detailed origin of such flares. We observe the signatures of `hardening' during the soft dips in the time variation of the dynamic photon index ($\Theta$) (Fig. 5(b)). \par 
     
          \item From the spectral analysis, we observe the normalized fluxes of the empirical \textsc{diskbb} and \textsc{power-law} model are comparable. The spectra can also be well fitted by invoking the reflection component coupled with the two-component model (TCAF). Therefore, the presence of two accretion flow components and the reflection of the Comptonized component from the disk seems to be well suited to explain the underlying flow properties. \par 
          \item In the case of $\nu$ class data of GRS 1915+105, the lightcurve is characterized by flaring activities, which is immediately followed by sharp dips and successive quasi-steady photon intensity over a few tens of seconds. The energy-resolved lightcurves in A, B and C bands show different patterns compared to IGR J17091-3624. While the C-band lightcurve of IGR J17091-3624 remained nearly non-variable, in the case of GRS 1915+105 the C-band intensity significantly diminishes during soft dips. It is also observed that the increase in intensity during flaring events is most prominent for A band (soft energy domain), while in C band any signature of increase in photon count is undetected (Fig. 2(a)). The intensity ratio plot as well as the normalized count plot corresponding to A, B and C bands also depict preferential decrease of C-band intensity during soft dips for GRS 1915+105 (Fig. 3(a) and 4(a)). This is a crucial difference between the timing properties of these two sources in their $\nu$ class. This strongly suggests that the $\nu$ class morphology of GRS 1915+105 may not be just because of the inter-conversion of the existing flow components in their brief persistent phase, but the disappearance of the additional power-law component originating out of the presence of the jet requires to be invoked to explain the observed spectro-temporal features.
          
          \item  Apart from the disk evacuation as a result of thermal-viscous instability or `magnetic rubber band' effect as envisaged, we propose the presence of micro-flare events and the additional Compton emission from the flares to explain the observed energy-resolved temporal features of GRS 1915+105. If a fraction of the seed photons is intercepted by the micro-flares, that could cause the seed photons to up-scatter and an additional power-law component would be produced. This phenomenon would lead to the cooling of the flares and subsequent feedback on the accreting material. This rapid feedback loop can be one possible mechanism to explain the observed temporal characteristics. Short duration intensity variation during the flaring events is indicative of the local accretion flow modulation triggered by this feedback loop of accreting material. This mechanism has been previously suggested in the case of $\chi$ and $\theta$ class variability as well (Banerjee et al. 2020a,b). Additional power-law component could also be attributed to the high energy tail of the synchrotron emission from the micro-ejections (Vadawale et al. 2001, 2003b).  \par 
          
            \item In the case of GRS 1915+105, we also detect sharp QPO around 7 Hz during the nearly steady phase of its lightcurve in all A, B and C bands. The radial oscillation of the shock front causes the Comptonized component to oscillate, which is manifested as the sharp QPO corresponding to the PDS. The oscillation of the Comptonizing region causes the additional power-law component to oscillate as well. Since the movement of the inner edge of the disk is governed by the viscous time-scale, it can not oscillate in tune with the shock boundary. Therefore, the fraction of the Comptonized photons within 2.0-6.0 keV range contributes to the A-band QPO. However, once the accretion flow modulation because of the local feedback loop happens, the cooling becomes more efficient which makes the cooling time scale go down. This leads to the disappearance of QPOs, as observed during the transition of the source from the quasi-variable phase to the flaring phase. \par 
     
    \item We report the presence of prominent QPO around 23 mHz in the case of IGR J17091-3624 in its $\nu$ class. We also examine the energy dependence of this QPO and observe that QPO is prominent in A and B bands, but any signature of QPO is not observed in C-band.  This lower frequency sharp QPO in A and B bands are indicative of the larger length scale of oscillation of the source in that particular variability class. The oscillation of the inner edge of the disk, as well as the Compton cloud, are attributed to this oscillation as before; however, lower QPO frequency is suggestive of higher shock location. Such energy dependence of QPO is also observed in C2 class, but the QPO happens to be broader in C2 class. We note, however, that the QPO frequency also depends on the mass of the black hole since the length scale scales with the mass. It will also depend on the strength of the shock, which, in turn, is dependent on the ratio of the two accretion rates as well as the dynamical parameters like viscosity. We, however, do not attempt to provide any conclusive physical justification regarding the difference of the QPO frequencies pertaining to these objects.
          \item The notion regarding the presence of a possible third component is further reinforced from the spectral fitting as well. In the case of $\nu$ class of GRS 1915+105, statistically acceptable spectral fits could not be achieved by adding the reflection component with TCAF only. We had to invoke an additional \textsc{cutoff power-law} model to incorporate the additional emission from the ejections as envisaged previously. However, during the spectral fitting of IGR J17091-3624, such an additional power-law was not necessary and the high energy Comptonized component remained persistent. This connection between the third additive component and the emission from the ejected material has earlier been drawn as well to explain the soft X-ray dips in GRS 1915+105 (Vadawale 2001a,b).
          
               \item Finally, the nearly steady C-band intensity is detected in the case of C2 class observation as well. Like $\nu$ class data of IGR J17091-3624, the normalized count in C-band remains nearly non-variable across the sharp dips and flares (Fig. 12). However, $\Theta > 0$ values are obtained during soft dips, implying quick and brief hardening of the source during the dips (Fig. 13(a,b)). The transition from variable to non-variable phase is not marked by any abrupt changes in $\Theta$, rather there is a gradual transition from rapidly fluctuating to quasi-variable $\Theta$ value. The weak positive correlation between the C-band intensity with $\Theta$ as well as the satisfactory spectral fits using \textsc{TCAF+pexrav} model are quite resembling with the $\nu$ class of IGR J17091-3624. Therefore, the existence of two components, the inner disk instability and the evacuation of the disk matter, like in the case of $\nu$ class, can be invoked in this case as well to explain the underlying accretion flow scenario.
 \end{enumerate}
 
 Long term and continuous observation across other variability classes of these two sources and their comparative spectro-temporal studies can reveal further crucial informations and enable us to gain a holistic understanding regarding the underlying accretion flow properties triggering all these classes. This is an extensive exercise and would be taken up in upcoming papers. \par 
 
 After intense activity in X-ray and radio for more than two decades, GRS 1915+105 has turned dim after July 2018 (Negoro et al. 2018) as revealed by MAXI/GSC observation. After March 2019, the X-ray intensity declined further (Homan et al. 2019; Rodriguez et al.
2019), suggesting the hypothesis that its 26yr of intense activity is nearing to an end. Renewed activity was observed for a brief period of $\sim$ 1 month (Balakrishnan et al. 2019, Miller et al. 2019), and the object has remained faint since. This obscured state of GRS 1915+105 is being monitored by multi-wavelength campaigns and would potentially reveal new evidences regarding the details of the source, which could be instrumental in future in further exploring this intriguing variability as well.

\section*{Acknowledgement}
We thank the anonumous referee for their comments and suggestions that have improved the manuscript. This research made use of the data obtained from HEASARC provided by the NASA archive (\url{https://heasarc.gsfc.nasa.gov/}). A. Banerjee acknowledge the fellowship support of the S. N. Bose National Centre for Basic Sciences, Kolkata,
India. A. Bhattacharjee's work was supported by the National Research Foundation (NRF) of Korea through grants 2016R1A5A1013277. DD and SKC acknowledge partial supports from the Department of Higher Education, Government of West Bengal, India and ISRO sponsored RESPOND project (ISRO/RES/2/418/17-18) fund.

\pagebreak

\begin{table} [h!]
{{\bf Table 1:} QPO parameters of the fundamental QPO frequency of GRS 1915+105 obtained from Lorentzian fitting.}
\centering

\begin{tabular}{  c c c c  }
\hline
Data  & $ v_{\text{QPO}}^b $ & $ \text{FWHM (Hz)}^c $ & $ \text{RMS}^d $  \\ [0.5ex]
$ \text{Specification}^a $ & (Hz) & & (\%)   \\ [0.5ex]
 \hline
$\text{full}^e$ & $7.19 \pm 0.04$ & $0.42 \pm 0.03$ & $4.55 \pm 0.13$ \\
A &  $7.19 \pm 0.03$ & $0.20 \pm 0.03$ & $3.27 \pm 0.11$ \\
B &  $7.19 \pm 0.05$ & $0.25 \pm 0.05$ & $5.24 \pm 0.16$ \\
C &  $7.20 \pm 0.04$ & $0.37 \pm 0.07$ & $9.01 \pm 0.18$ \\ [1ex]
\hline
\end{tabular}
\caption*{Note: $^a$ The QPO parameters in different energy bands.  $^b$ The centroid frequency of QPO. $^c$ Full width at half maxima. $^d$ The fractional rms power corresponding to QPO. $^e$ corresponds to 2.0-60.0 keV.}
\end{table}

\begin{center}
{{\bf Table 2:} Best-fit parameters obtained from the spectral fits corresponding to the $\nu$ class of GRS 1915+105 and IGR J17091-3624.}
    \begin{table}[h!]
     
  \begin{tabular}{l l l l l}
    \hline
    \vspace{0.15cm}
    \multirow{2}{*}{Model parameters} &
     
      \multicolumn{2}{c}{GRS 1915+105} &
      \multicolumn{2}{c}{IGR J17091-3624} \\
    & \textsc{tcaf+cutoffpl} & \textsc{diskbb+nthcomp} & \textsc{tcaf+pexrav} & \textsc{diskbb+power-law}   \\
    & \textsc{+pexrav} &  \textsc{+gaussian} &  & \textsc{+gaussian} \\
        \hline
    \vspace{0.2cm}
 $T_{\text{in}}$ & -- & $1.20_{-0.04}^{+0.03}$ & -- & $1.11_{-0.05}^{+0.04}$\\
         \vspace{0.2cm}
$\Gamma_{\text{po}}$ & -- & -- & -- & $2.15_{-0.04}^{+0.05}$\\
         \vspace{0.2cm}

 $\Gamma_{\text{nth}}$ & -- & $2.69_{-0.10}^{+0.12}$ & -- & --\\
         \vspace{0.2cm}

 $kT_e$ & -- & $24.33_{-1.25}^{+1.10}$ & -- & --\\
 \hline
        \vspace{0.2cm}

  $\dot{m}_d$ & $1.02_{-0.07}^{+0.06}$ & -- & $0.12_{-0.01}^{+0.01}$  & -- \\
    \vspace{0.2cm}
    $\dot{m}_h$ & $0.30_{-0.03}^{+0.03}$ & -- & $0.25_{-0.02}^{+0.01}$ & -- \\
        \vspace{0.2cm}

    $X_s$ & $19.68_{-1.40}^{+1.20}$ & -- & $35.41_{-1.70}^{+1.64}$ & -- \\
        \vspace{0.2cm}

    $R$ & $1.09_{-0.03}^{+0.02}$ & -- & $2.21_{-0.03}^{+0.04}$ & --\\
\hline
    \vspace{0.2cm}

    $\alpha$ & $0.66_{-0.02}^{+0.03}$ & -- & -- & -- \\
     \vspace{0.2cm}
    
    $\beta$ & $2.28_{-0.13}^{+0.12}$ & -- & -- & --\\
    \hline
    \vspace{0.2cm}
   $\Gamma_{\text{refl}}$ & $1.09_{-0.03}^{+0.04}$ & -- & $1.10_{-0.04}^{+0.03}$ & --\\
 \vspace{0.2cm}
 
    $E_c$ & $8.31_{-0.08}^{+0.09}$ & -- & $1.79_{-0.02}^{+0.04}$ & -- \\
    \vspace{0.2cm}
    
    $\text{rel}_\text{refl}$ & $2.34_{-0.07}^{+0.08}$ & -- & $5.77_{-0.08}^{+0.06}$ & --\\
     \hline
    $\chi^2_\text{DOF}$ & $31.31/41$ & $41.30/43$ &  $53.37/43$  &  $47.92/43$\\
    \hline
  \end{tabular}
  \caption*{Note: $T_{\text{in}}$: inner disk temperature in keV. $\Gamma_{\text{po}}$: photon index from \textsc{diskbb} model. $\Gamma_{\text{nth}}$: photon index from \textsc{nthcomp} model. $kT_{\text{in}}$: electron temperature from \textsc{nthcomp} model in keV. $\dot{m}_d$: disk rate in Eddington unit. $\dot{m}_h$: halo rate in Eddington unit. $X_s$: Shock location in $r_g$. $R$: Shock strength. $\alpha$: power law photon index. $\beta$: e-folding energy of exponential rolloff (in keV). $\Gamma_{\text{refl}}$: photon index from \textsc{pexrav}. $E_c$: cutoff energy from \textsc{pexrav} in keV. $\text{rel}_\text{refl}$: reflection fraction. In \textsc{nthcomp}, the seed photon temperature has been tied to $T_{\text{in}}$.}
\end{table}
\end{center}

\clearpage

\begin{figure}
     \centering
     \begin{subfigure}[b]{0.48\textwidth}
         \centering
         \includegraphics[width=\textwidth]{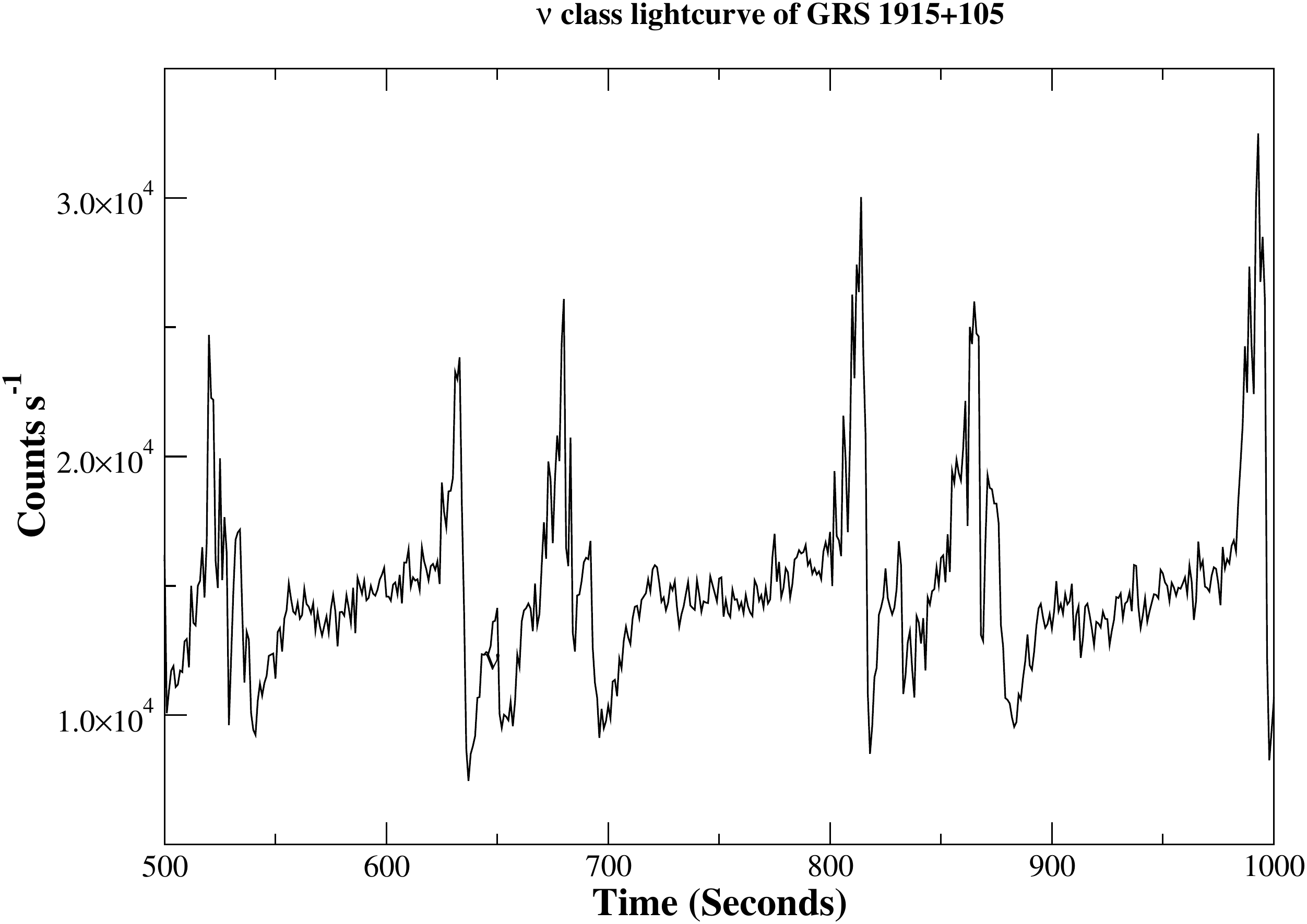}
         \caption{}
     \end{subfigure}
     \hfill
     \begin{subfigure}[b]{0.48\textwidth}
         \centering
         \includegraphics[width=\textwidth]{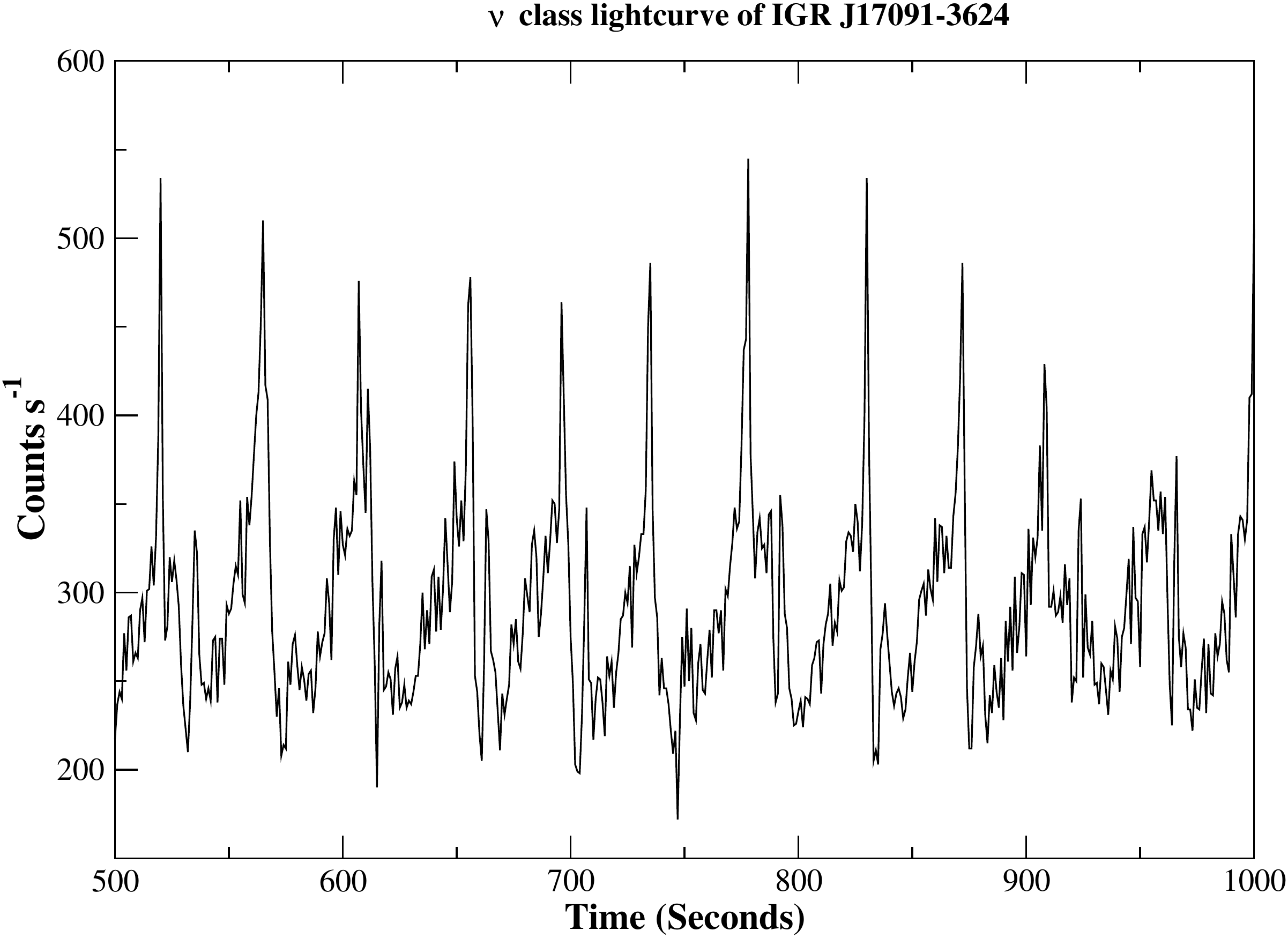}
         \caption{}
     \end{subfigure}
     \hfill
\caption{Comparison of lightcurves of (a) GRS 1915+105 and (b) IGR J17091-3624 in their $\nu$ class. The time-resolution of the lightcurves are 1.0 Sec. The recurrence time of micro flares is shorter in the case of IGR J17091-3624. In case of GRS 1915+105 the photon count remains steady for $\sim 50$ sec. in between the flares, but in IGR J17091-3624 such steady phase is not detected.}
\end{figure}

\begin{figure}
     \centering
     \begin{subfigure}[b]{0.48\textwidth}
         \centering
         \includegraphics[width=\textwidth]{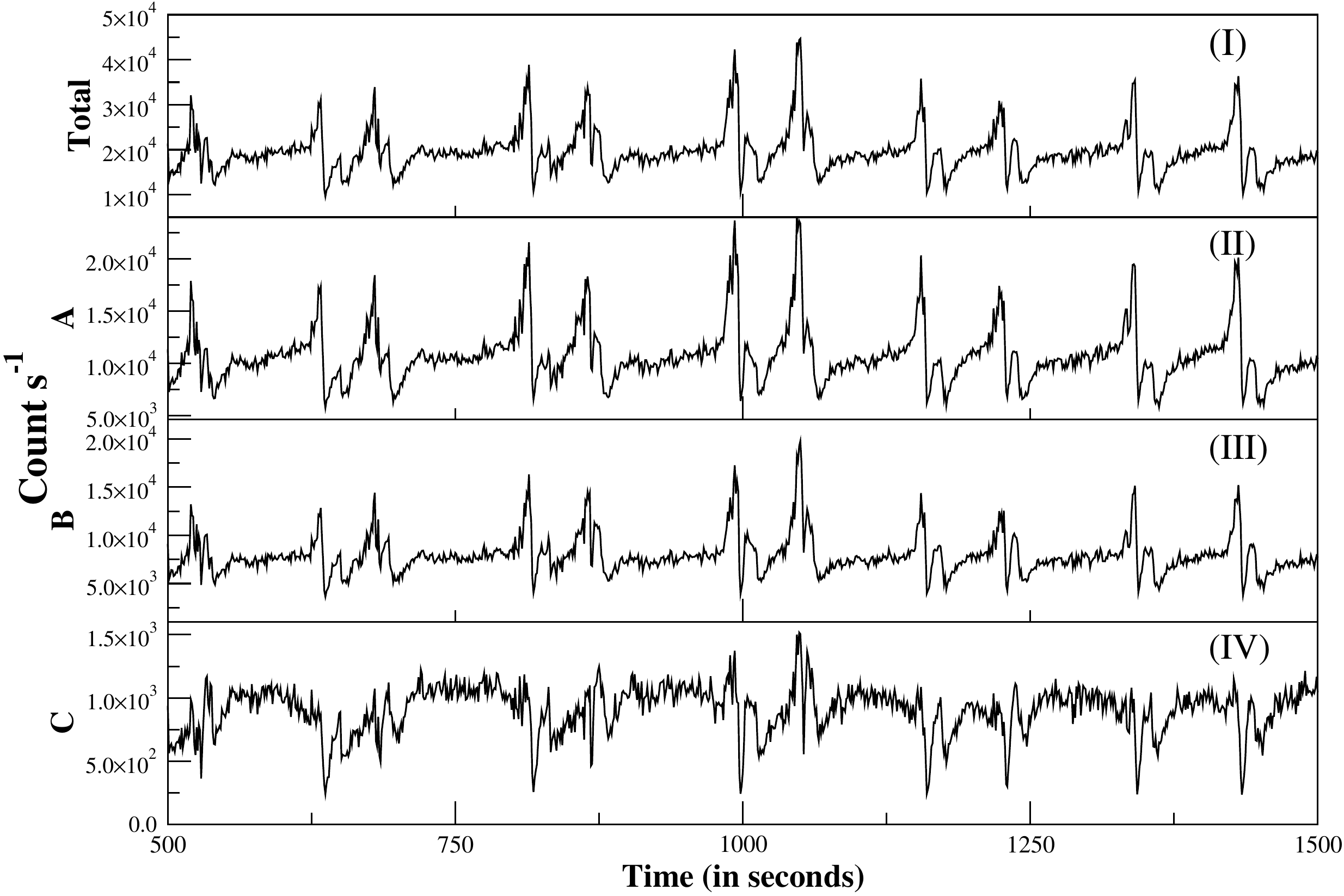}
         \caption{}
     \end{subfigure}
     \hfill
     \begin{subfigure}[b]{0.48\textwidth}
         \centering
         \includegraphics[width=\textwidth]{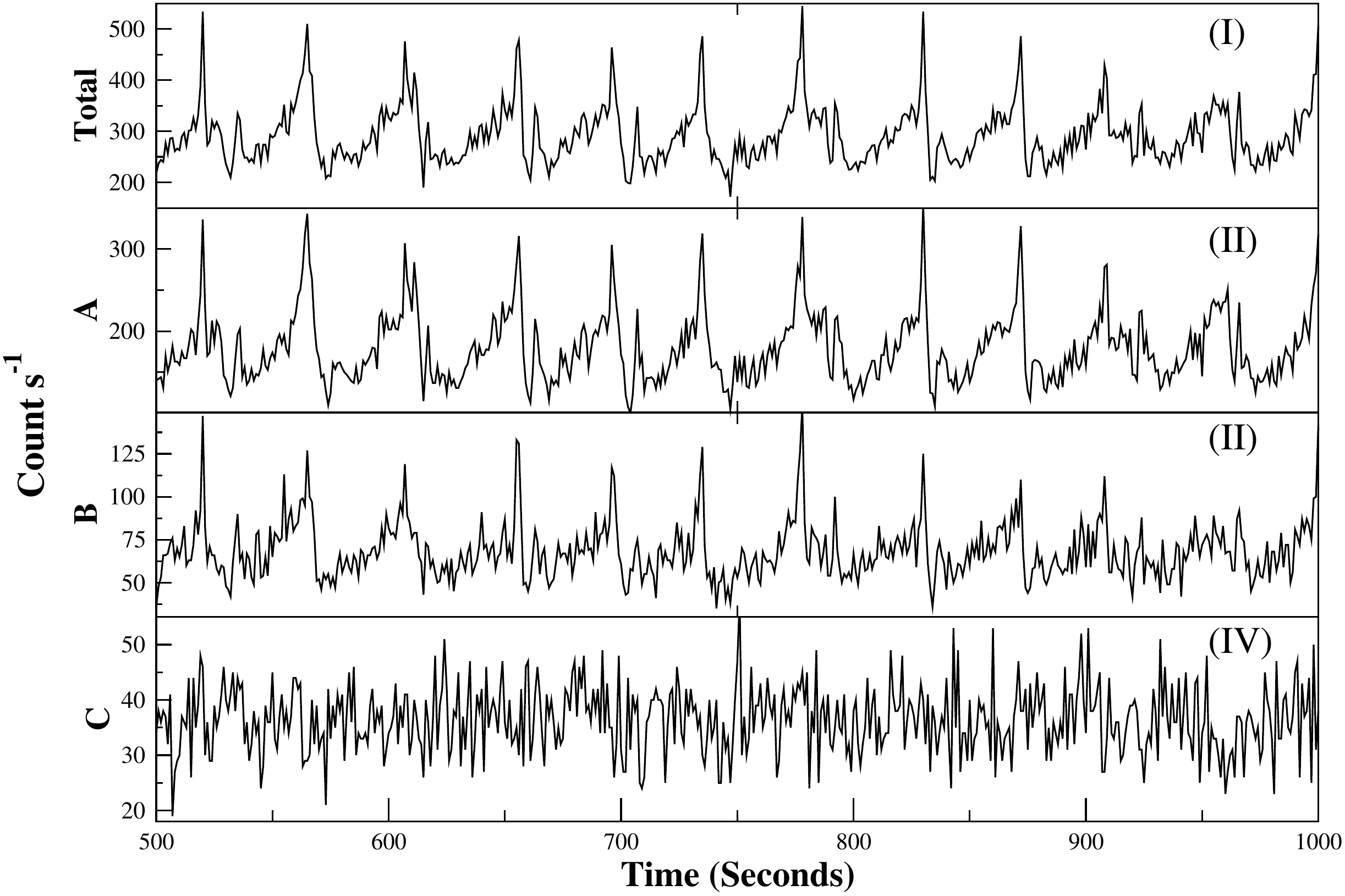}
         \caption{}
     \end{subfigure}
     \hfill
\caption{(a) Panel (I) contains broadband (2.0 -60.0 keV) lightcurve of GRS 1915+105 in its $\nu$ class. The time-resolution of the lightcurves are 1.0 Sec. In panels (II-IV), the lightcurves pertaining to A (2.0 -6.0 keV), B (6.0 - 15.0 keV) and C (15.0 - 60.0 keV) are provided. 2(b) Same sequence of lightcurves corresponding to IGR J17091-3624.}
\end{figure}

\begin{figure}
     \centering
     \begin{subfigure}[b]{0.48\textwidth}
         \centering
         \includegraphics[width=\textwidth]{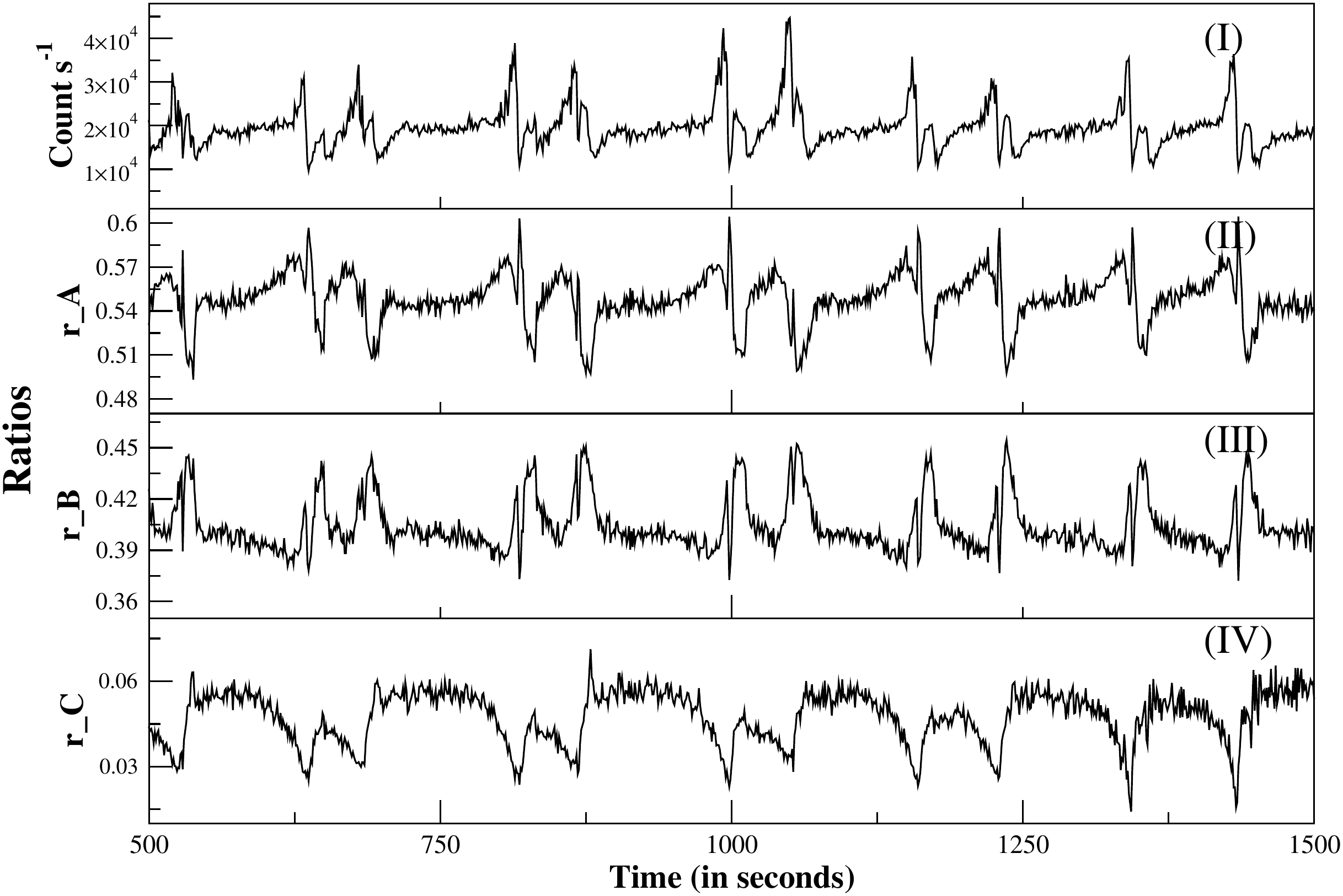}
         \caption{}
     \end{subfigure}
     \hfill
     \begin{subfigure}[b]{0.48\textwidth}
         \centering
         \includegraphics[width=\textwidth]{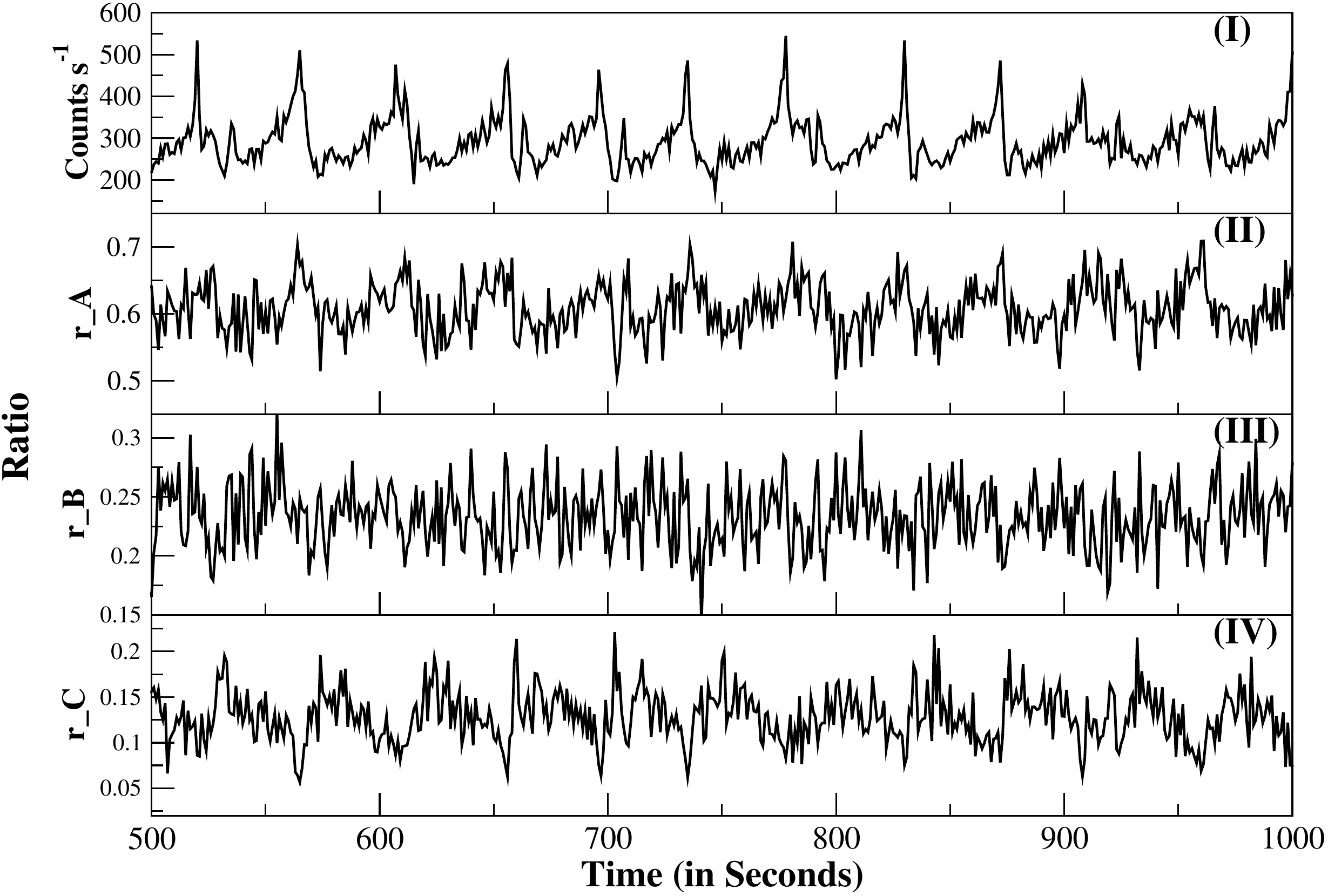}
         \caption{}
     \end{subfigure}
     \hfill
\caption{(a) The panel (I) contains the broadband lightcurve (2.0 -60.0 keV) of GRS 1915+105. In panels (II-IV), the ratios of A (2.0 - 6.0 keV), B (6.0 - 15.0 keV) and C (15.0 - 60.0 keV) band intensities with the broadband intensity (i.e. $r\_A, r\_B$ and $r\_C$ respectively) are plotted. 3(b) Same sequence of broadband lightcurves and intensity ratios are plotted corresponding to IGR J17091-3624.}
\end{figure}

\begin{figure}
     \centering
     \begin{subfigure}[b]{0.48\textwidth}
         \centering
         \includegraphics[width=\textwidth]{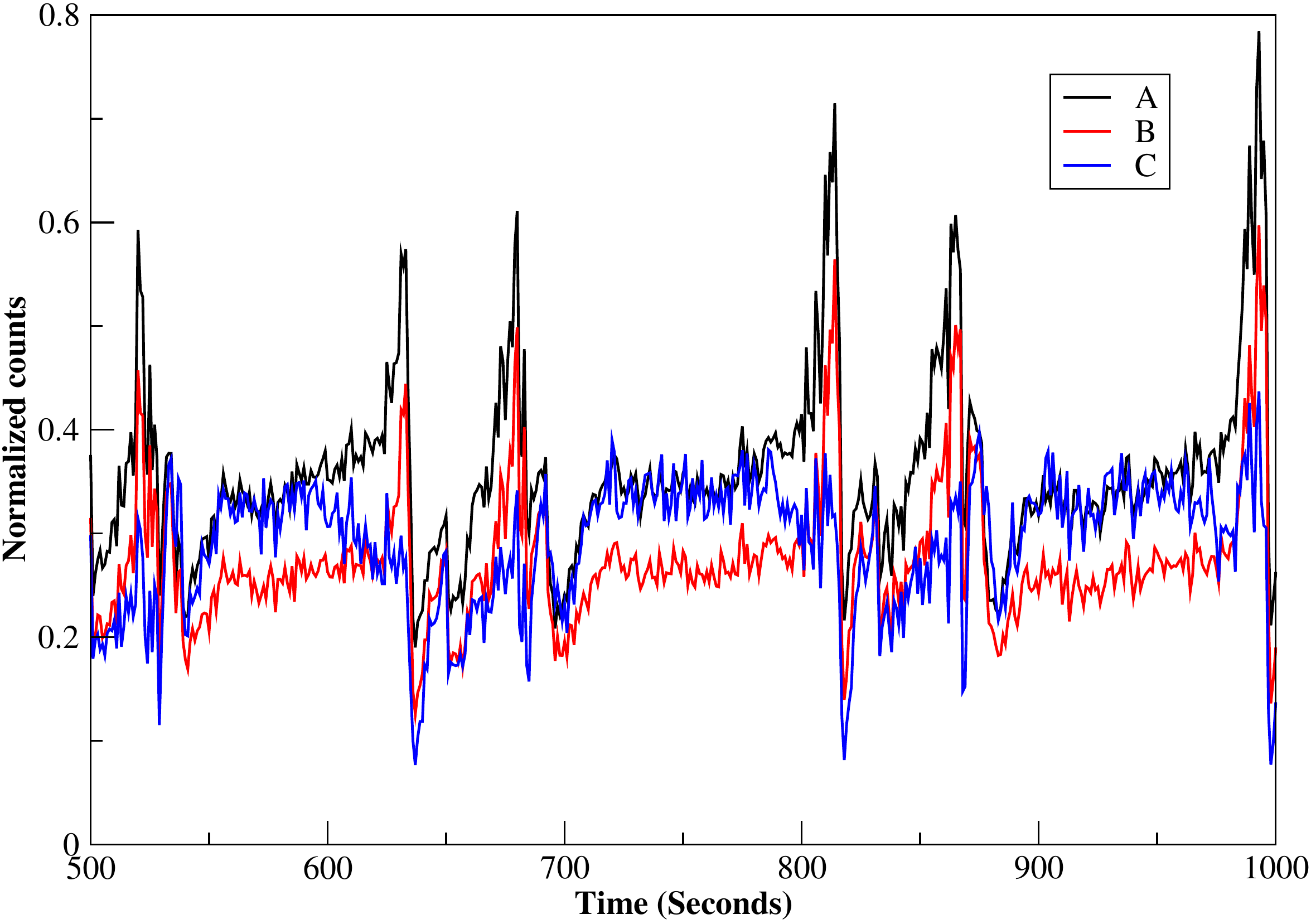}
         \caption{}
     \end{subfigure}
     \hfill
     \begin{subfigure}[b]{0.48\textwidth}
         \centering
         \includegraphics[width=\textwidth]{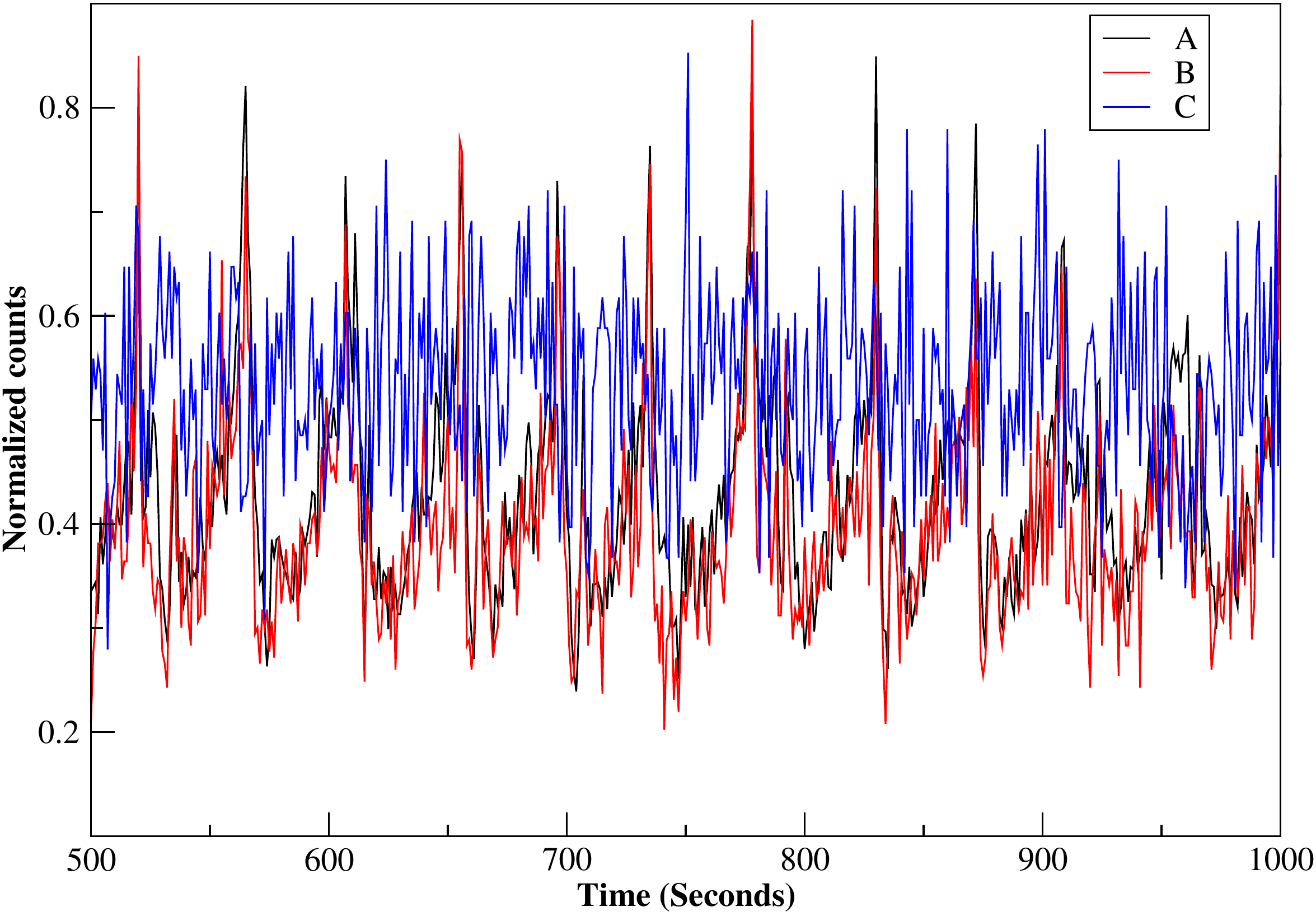}
         \caption{}
     \end{subfigure}
     \hfill
\caption{(a) The normalized counts of GRS 1915+105 in A (2.0 - 6.0 keV), B (6.0 - 15.0 keV) and C (15.0 - 60.0 keV) band are plotted together. We observe the preferential decrease in C-band count during the sharp intensity dips. During the flares, C band intensity does not increase at all. 4(b) Same normalized counts for IGR J17091-3624. We observe quasi variable C band intensity across the dips and flares.}
\end{figure}

\begin{figure}
     \centering
     \begin{subfigure}[b]{0.48\textwidth}
         \centering
         \includegraphics[width=\textwidth]{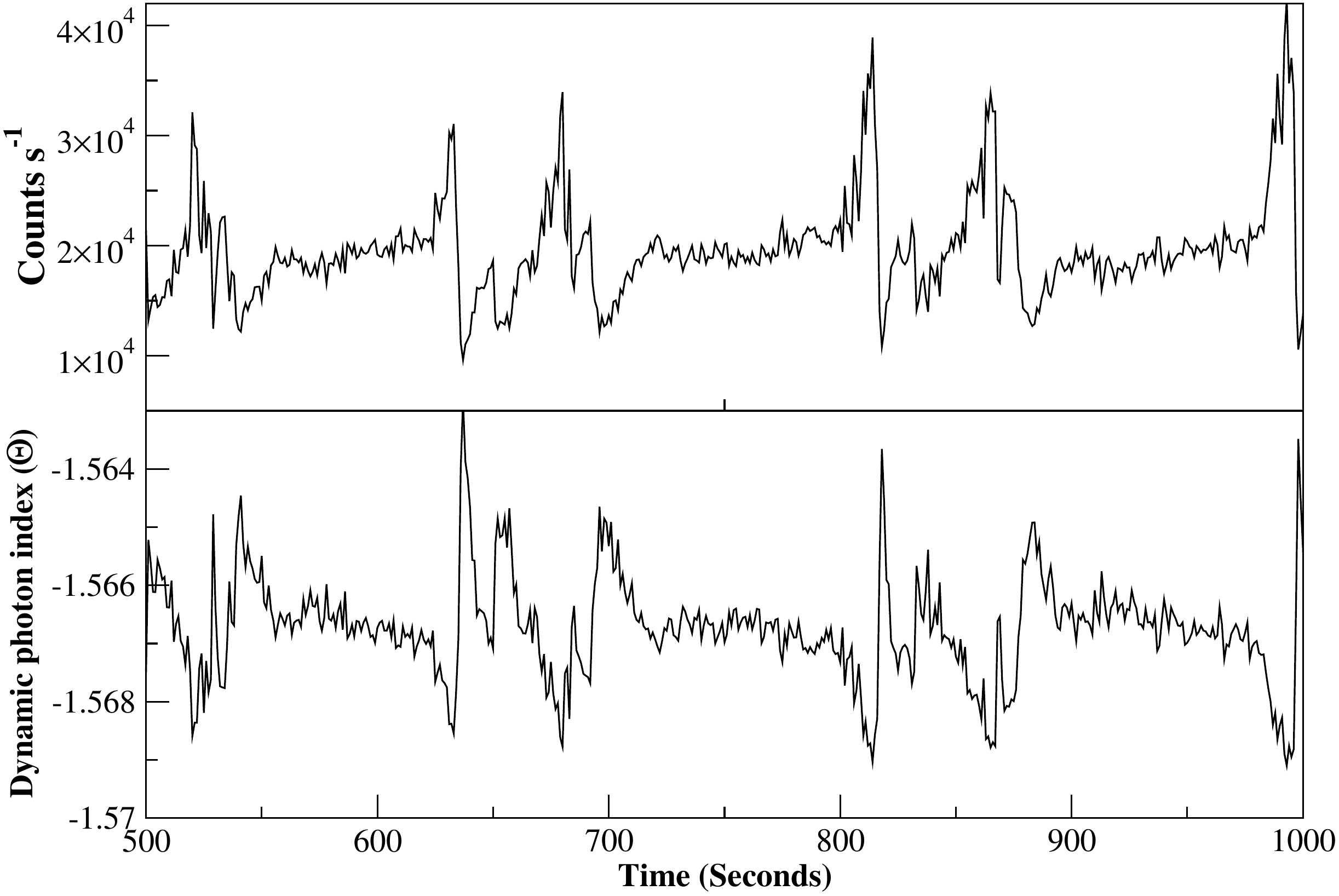}
         \caption{}
     \end{subfigure}
     \hfill
     \begin{subfigure}[b]{0.48\textwidth}
         \centering
         \includegraphics[width=\textwidth]{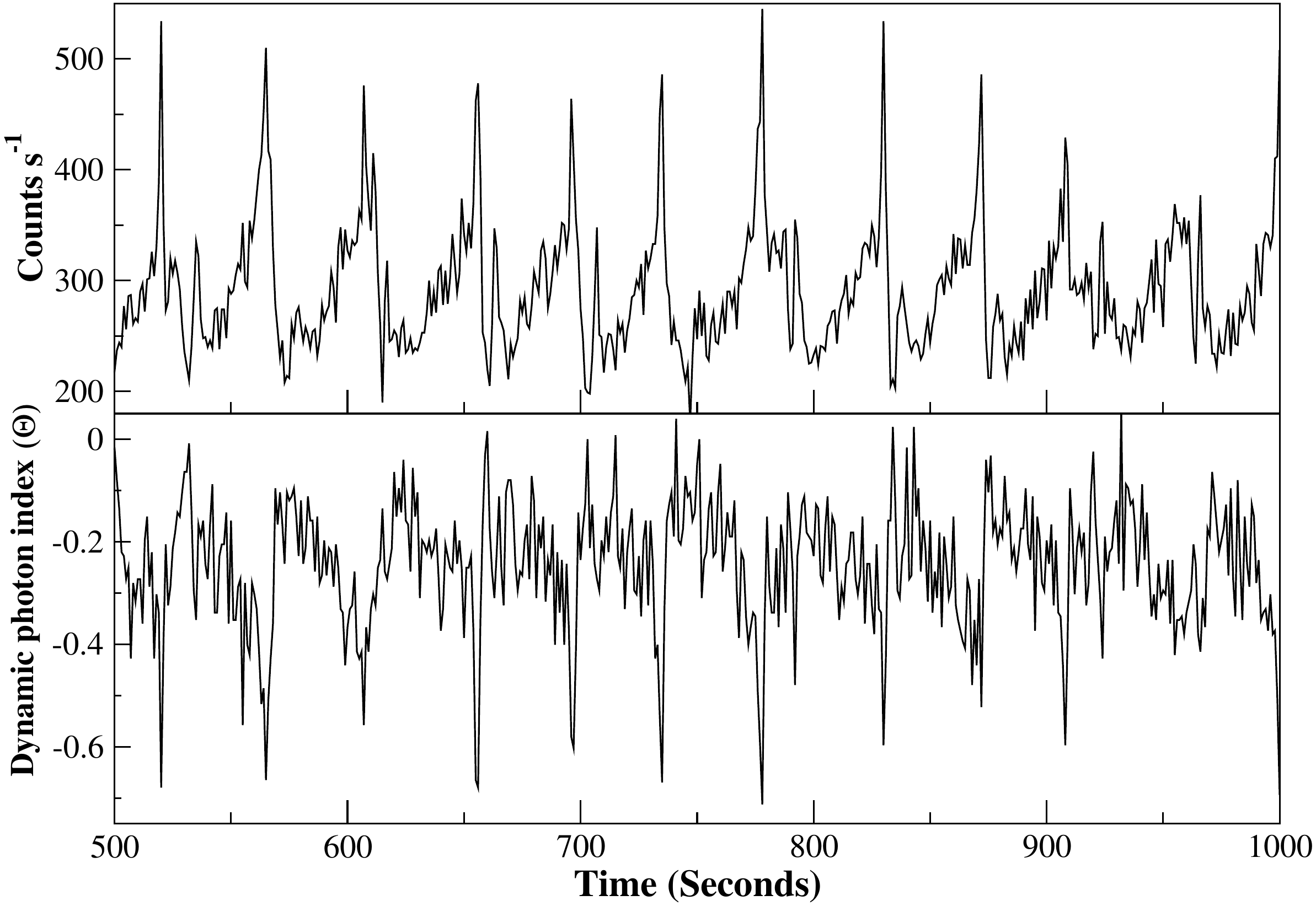}
         \caption{}
     \end{subfigure}
     \hfill
\caption{(a) The top and the bottom panels contain the variation of broadband intensity (2.0 - 60.0 keV) and the variation of dynamic photon index ($\Theta$) respectively for GRS 1915+105 in its $\nu$ class. 5(b) Same sequence of plots for IGR J17091-3624. In this case, $\Theta$ $\sim 0$ during the dips.}
\end{figure}

\begin{figure}
     \centering
     \begin{subfigure}[b]{0.45\textwidth}
         \centering
         \includegraphics[width=\textwidth]{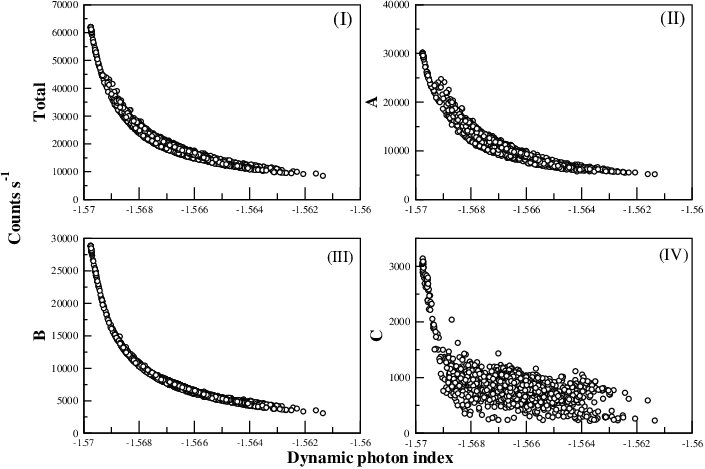}
         \caption{}
     \end{subfigure}
     \hfill
     \begin{subfigure}[b]{0.45\textwidth}
         \centering
         \includegraphics[width=\textwidth]{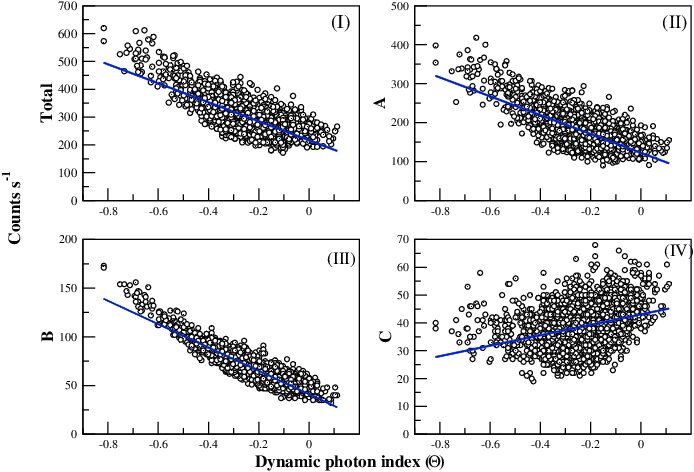}
         \caption{}
     \end{subfigure}
     \hfill
\caption{(a) Variation of $\Theta$ with broadband intensity (2.0 keV - 60.0 keV) as well as A (2.0 - 6.0 keV), B (6.0 - 15.0 keV) and C (15.0 - 60.0 keV) intensities are plotted in panels (I-IV) respectively. (b) Same sequence of plots for IGR J17091-3624. In this case weak positive correlation is observed for C-band intensity.}
\end{figure}

\begin{figure}
     \centering
     \begin{subfigure}[b]{0.48\textwidth}
         \centering
         \includegraphics[width=\textwidth]{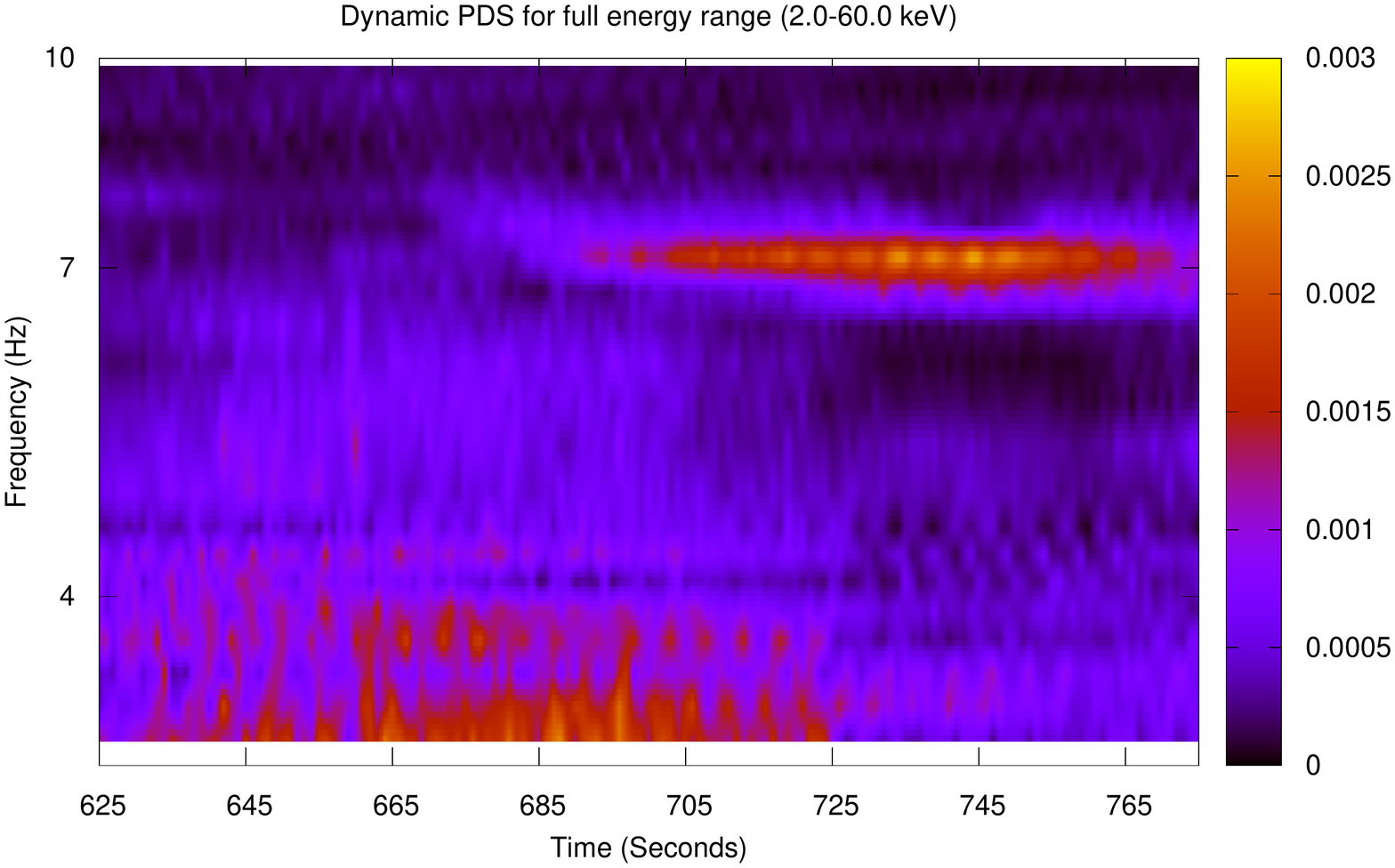}
         \caption{}
     \end{subfigure}
     \hfill
     \begin{subfigure}[b]{0.48\textwidth}
         \centering
         \includegraphics[width=\textwidth]{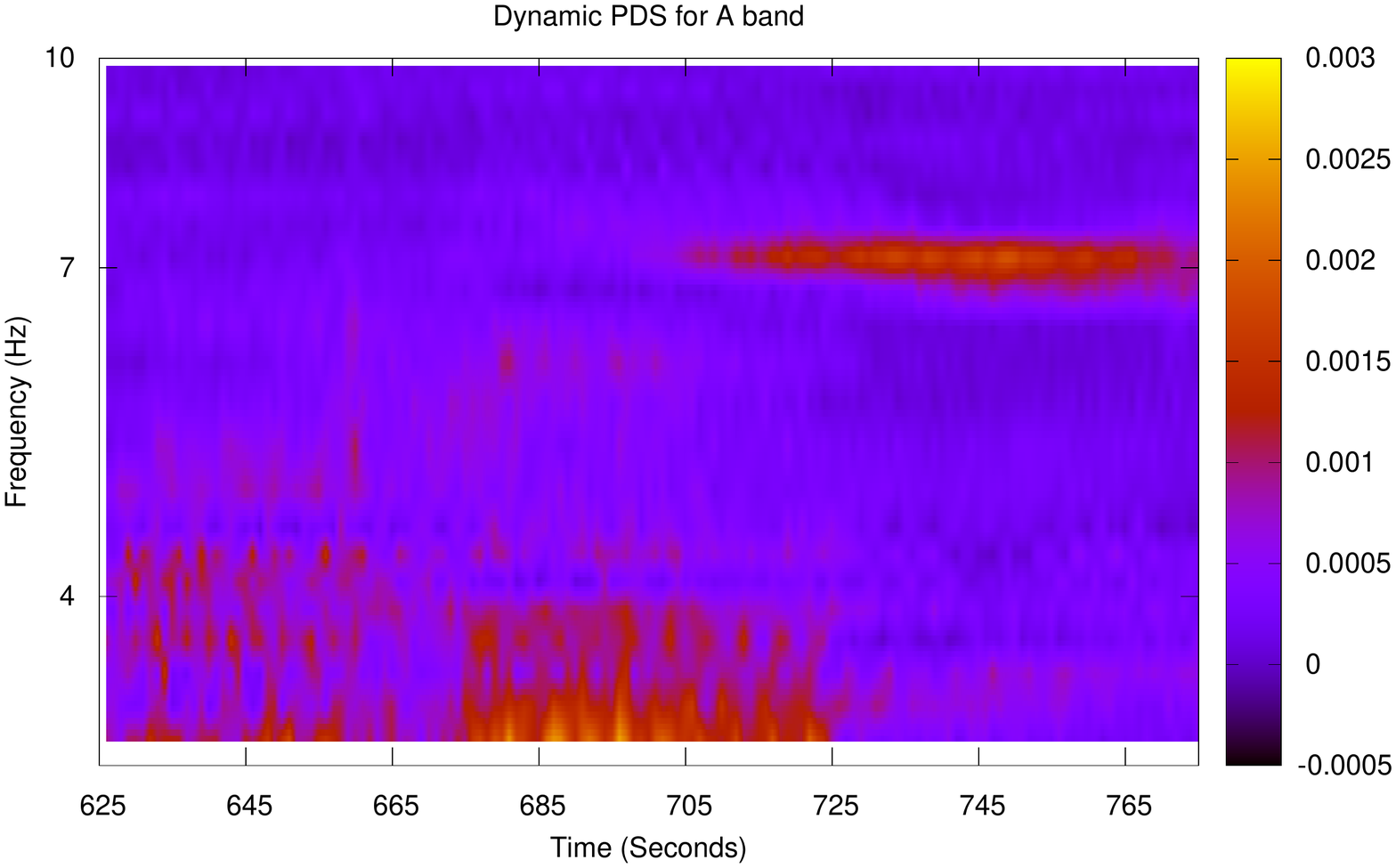}
         \caption{}
     \end{subfigure}
     \hfill
     \begin{subfigure}[b]{0.48\textwidth}
         \centering
         \includegraphics[width=\textwidth]{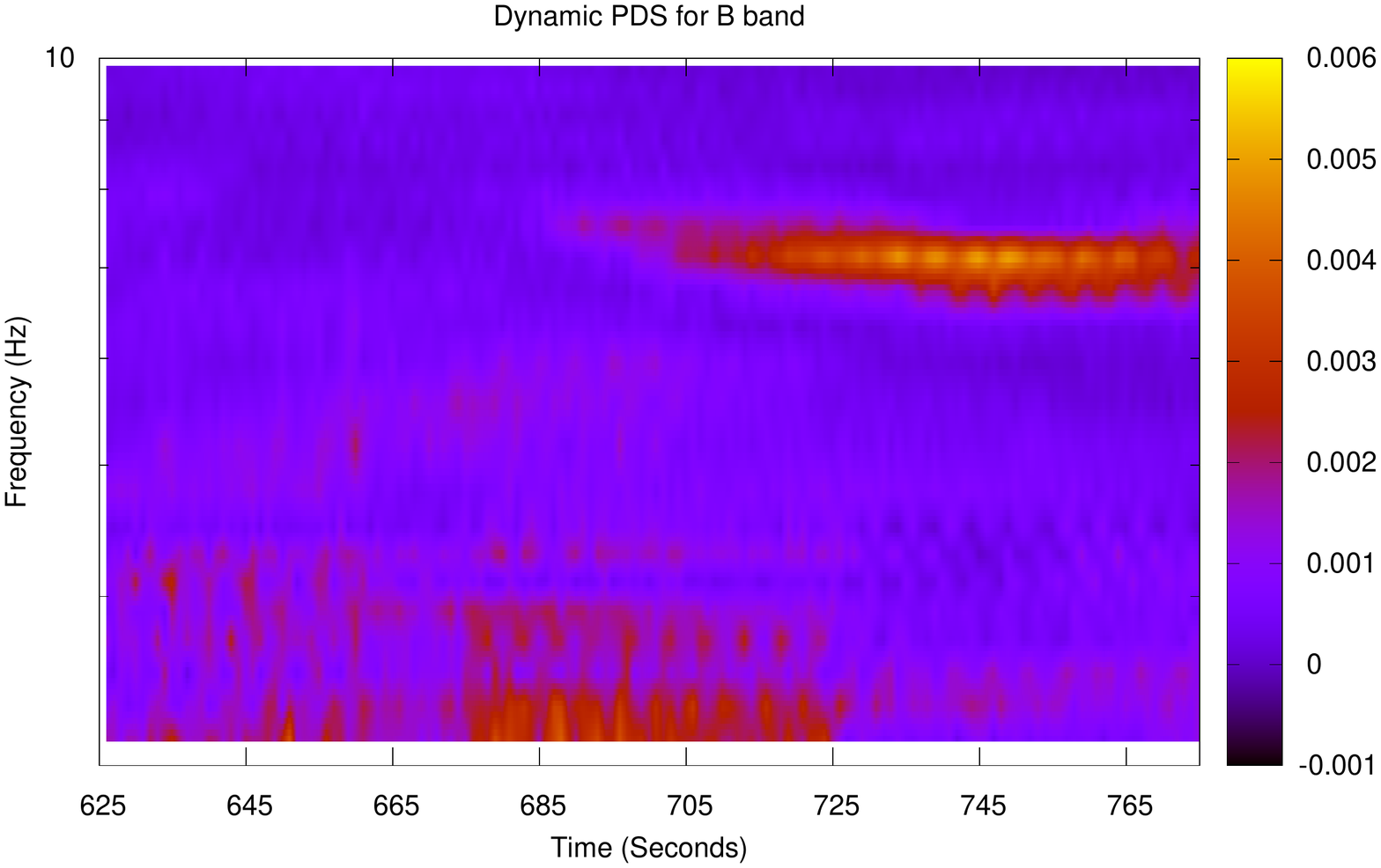}
         \caption{}
     \end{subfigure}
     \hfill
     \begin{subfigure}[b]{0.48\textwidth}
         \centering
         \includegraphics[width=\textwidth]{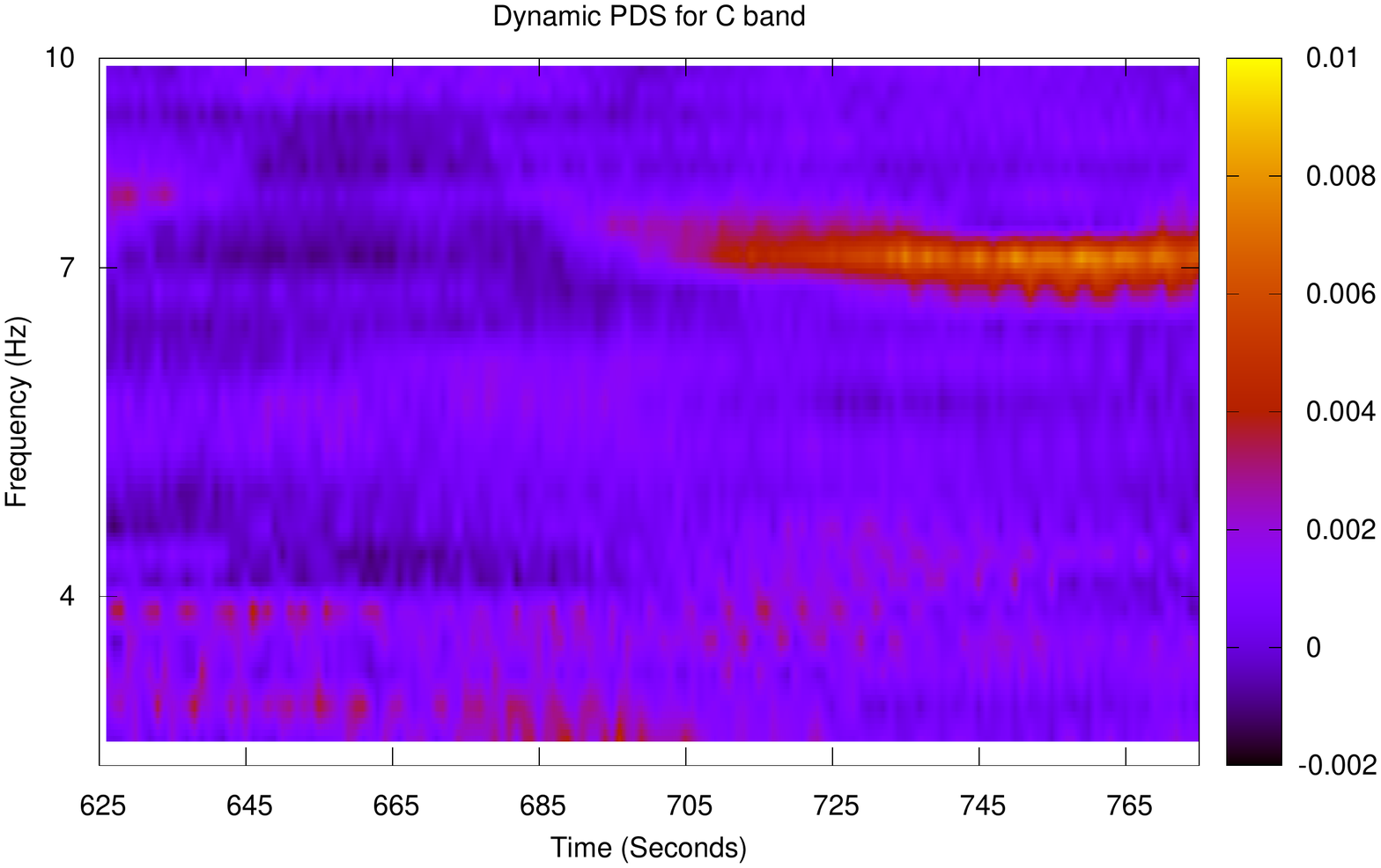}
         \caption{}
     \end{subfigure}
     \hfill
\caption{Dynamic power-density spectra for quasi-steady phase of GRS 1915+105 $\nu$ class data. Panel (a) contains dynamic PDS for broadband intensity (2.0 keV - 60.0 keV) and (b-d) contains dynamic PDS for A (2.0 - 6.0 keV), B (6.0 - 15.0 keV) and C (15.0 - 60.0 keV) band respectively.}
\end{figure}

\begin{figure}
     \centering
     \begin{subfigure}[b]{0.48\textwidth}
         \centering
         \includegraphics[width=0.85\textwidth,angle=-90]{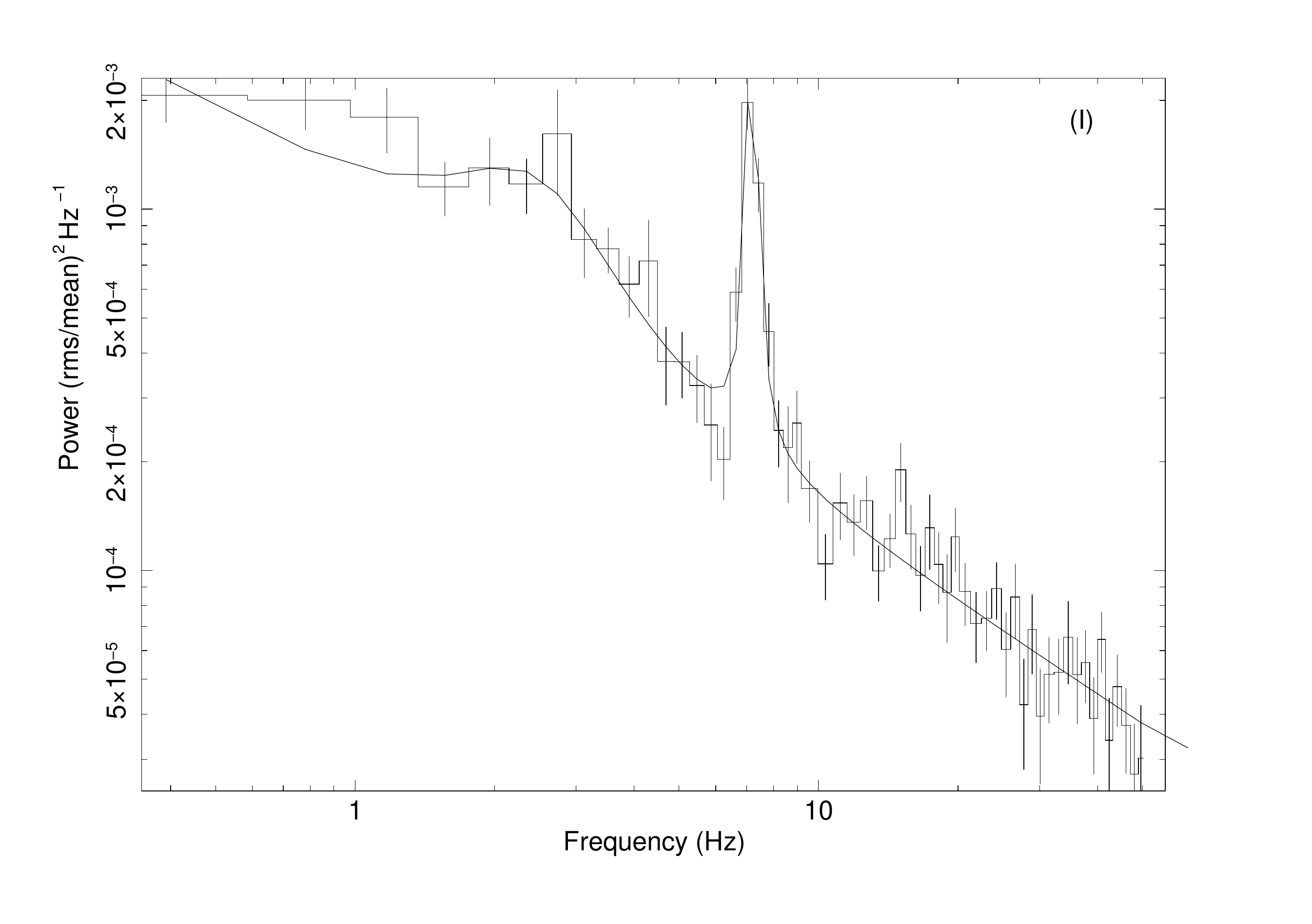}
         \caption{}
     \end{subfigure}
     \hfill
     \begin{subfigure}[b]{0.48\textwidth}
         \centering
         \includegraphics[width=0.85\textwidth,angle=-90]{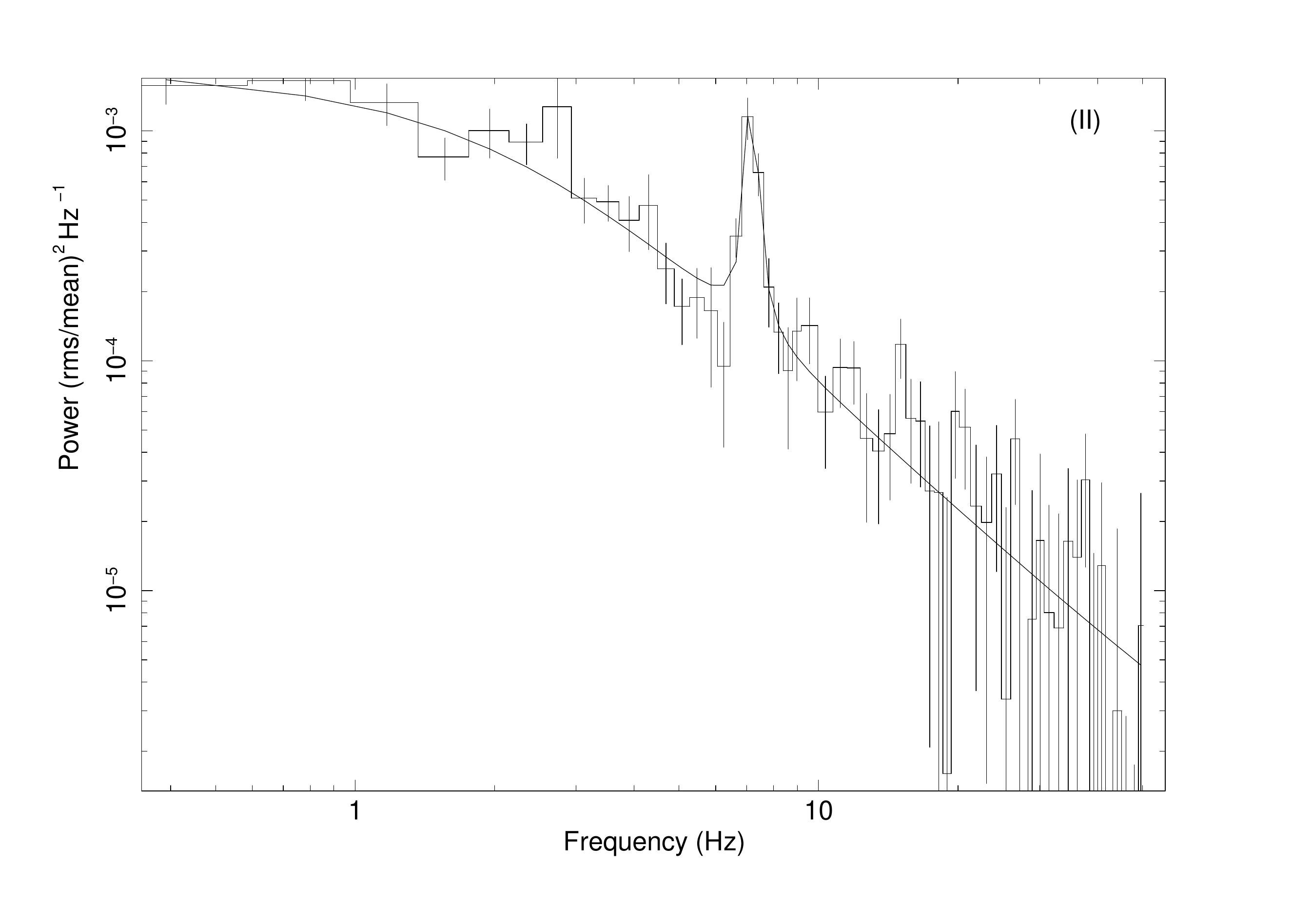}
         \caption{}
     \end{subfigure}
     \hfill
     \begin{subfigure}[b]{0.48\textwidth}
         \centering
         \includegraphics[width=0.85\textwidth,angle=-90]{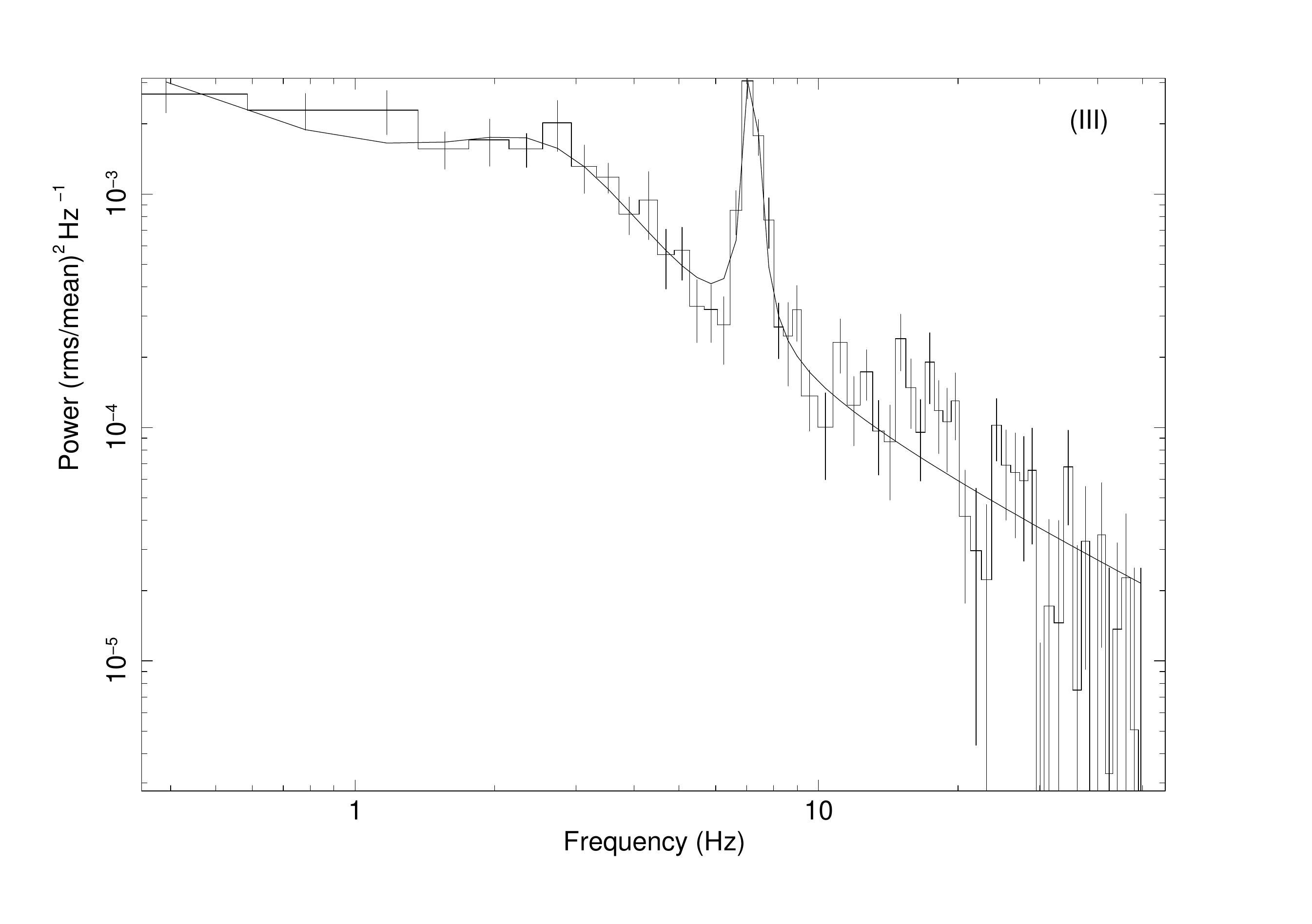}
         \caption{}
     \end{subfigure}
     \hfill
     \begin{subfigure}[b]{0.48\textwidth}
         \centering
         \includegraphics[width=0.85\textwidth,angle=-90]{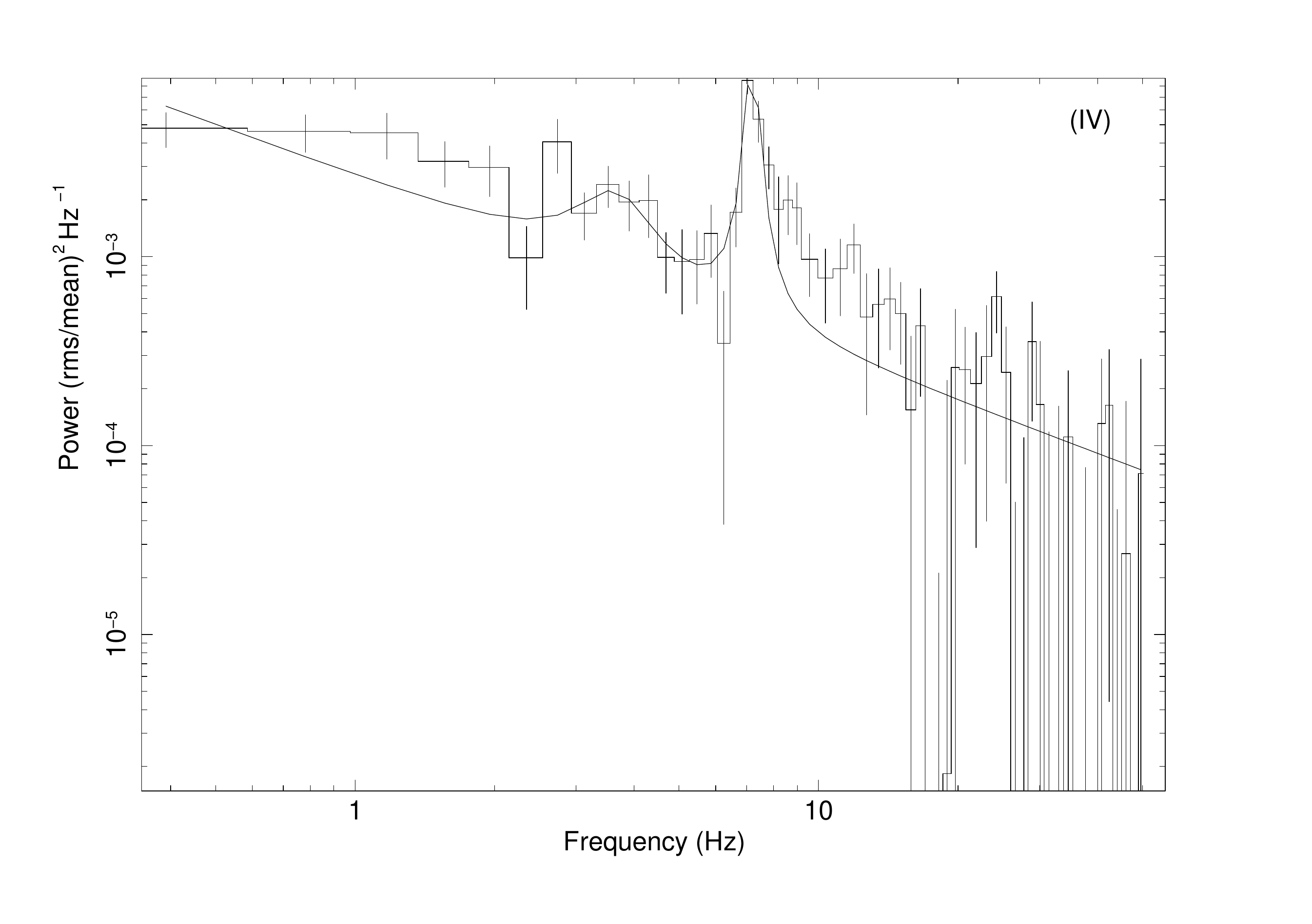}
         \caption{}
     \end{subfigure}
     \hfill
\caption{Power density spectra of GRS 1915+105, fitted with two Lorentzians and one power-law profile. Panel (I) contains PDS fit corresponding to the full 2.0 - 60.0 keV energy range, and panels (II-IV) contain PDS fit for A, B and C band. In all cases, sharp QPOs $\sim 7.1$ Hz are detected.}
\end{figure}

\begin{figure}
     \centering
     \begin{subfigure}[b]{0.48\textwidth}
         \centering
         \includegraphics[width=0.85\textwidth,angle=-90]{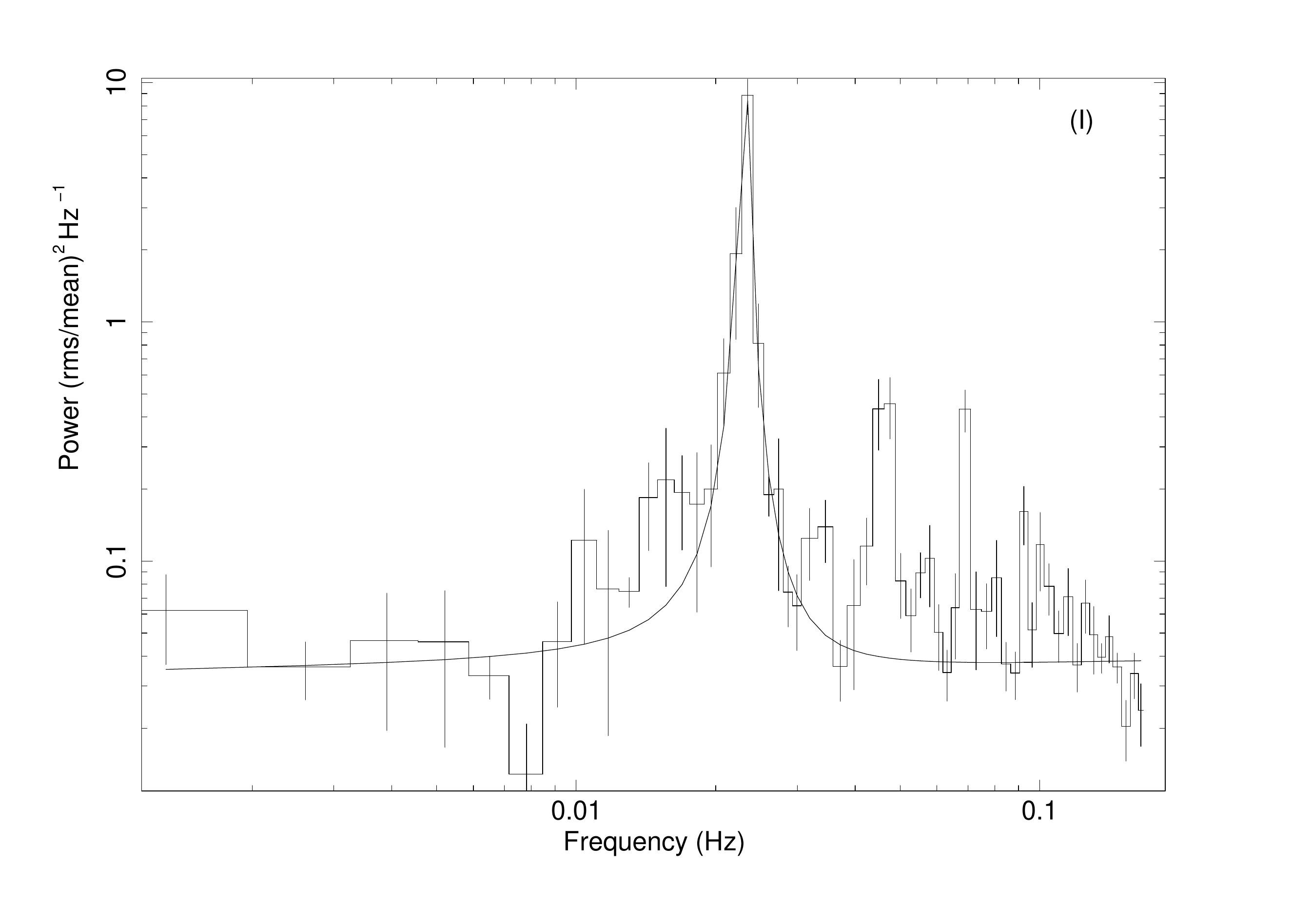}
         \caption{}
     \end{subfigure}
     \hfill
     \begin{subfigure}[b]{0.48\textwidth}
         \centering
         \includegraphics[width=0.85\textwidth,angle=-90]{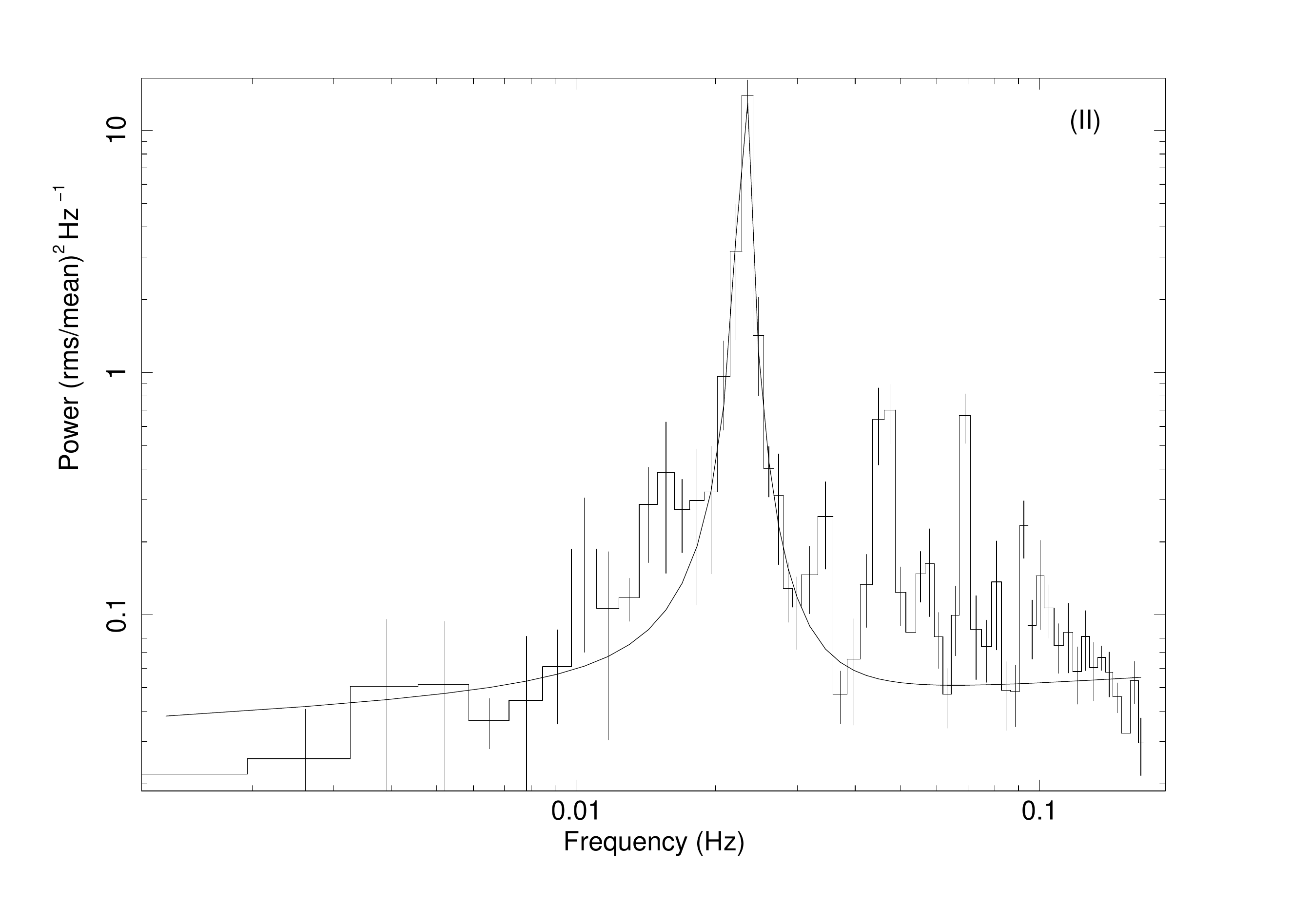}
         \caption{}
     \end{subfigure}
     \hfill
     \begin{subfigure}[b]{0.48\textwidth}
         \centering
         \includegraphics[width=0.85\textwidth,angle=-90]{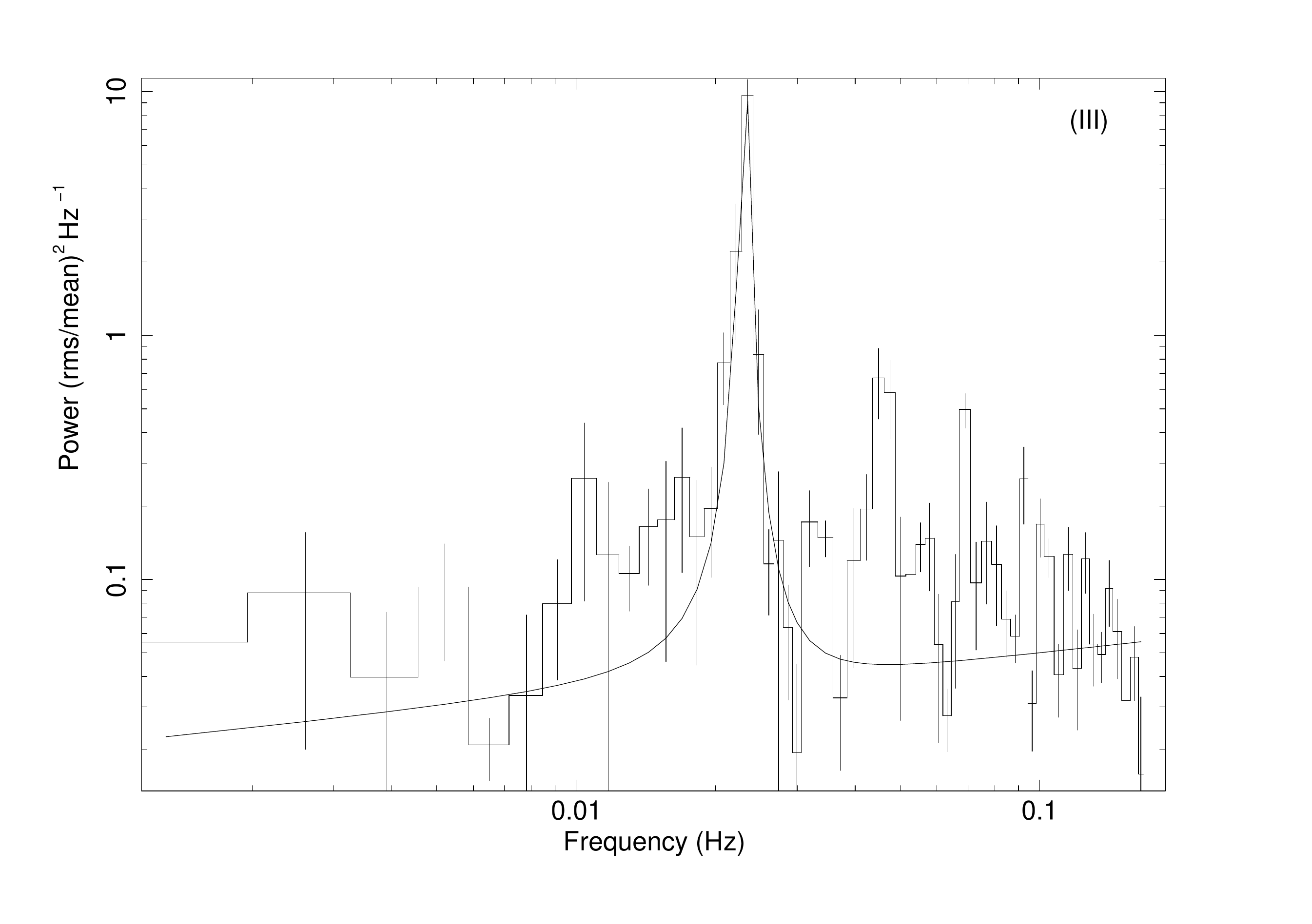}
         \caption{}
     \end{subfigure}
     \hfill
     \begin{subfigure}[b]{0.48\textwidth}
         \centering
         \includegraphics[width=0.85\textwidth,angle=-90]{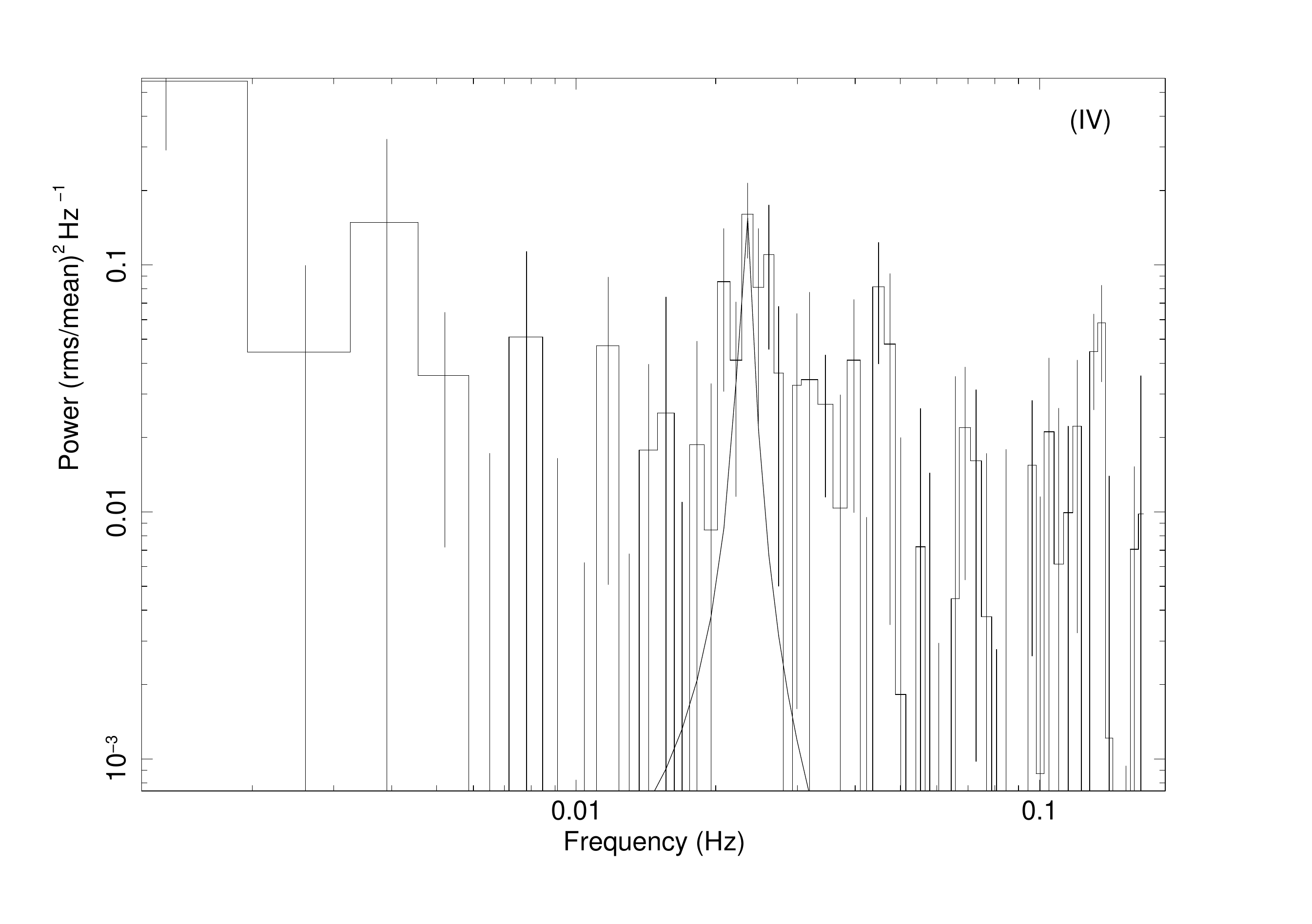}
         \caption{}
     \end{subfigure}
     \hfill
\caption{The Power density spectra of IGR J17091-3624 fitted with Lorentzian profile. Panel (I) contains PDS fit corresponding to the full 2.0 - 60.0 keV energy range, and panels (II-IV) contain PDS fit for A, B and C band. In all cases, sharp QPOs $\sim 23$ mHz are detected.}
\end{figure}

\begin{figure}
     \centering
     \begin{subfigure}[b]{0.48\textwidth}
         \centering
         \includegraphics[width=\textwidth]{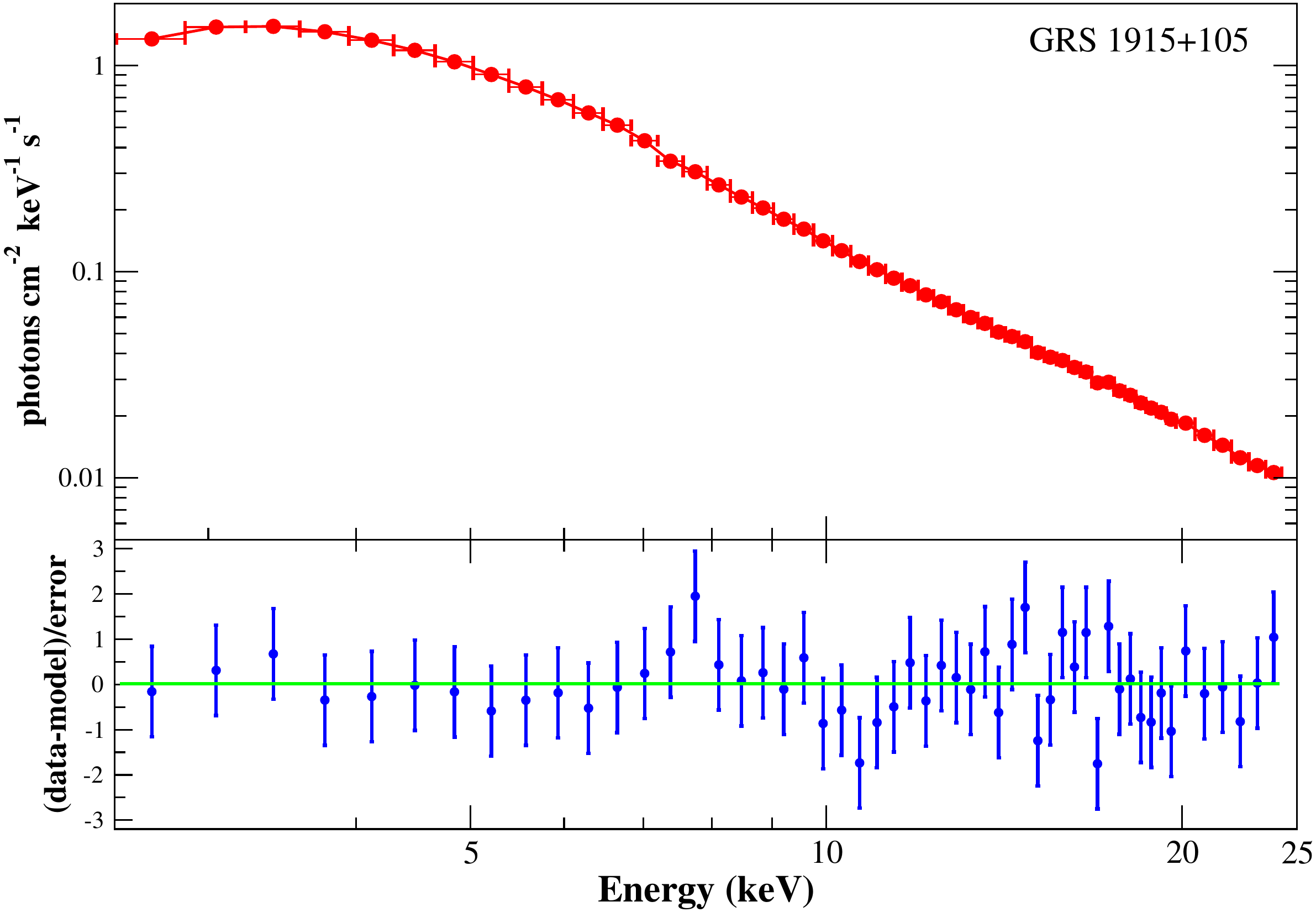}
         \caption{}
     \end{subfigure}
     \hfill
     \begin{subfigure}[b]{0.48\textwidth}
         \centering
         \includegraphics[width=\textwidth]{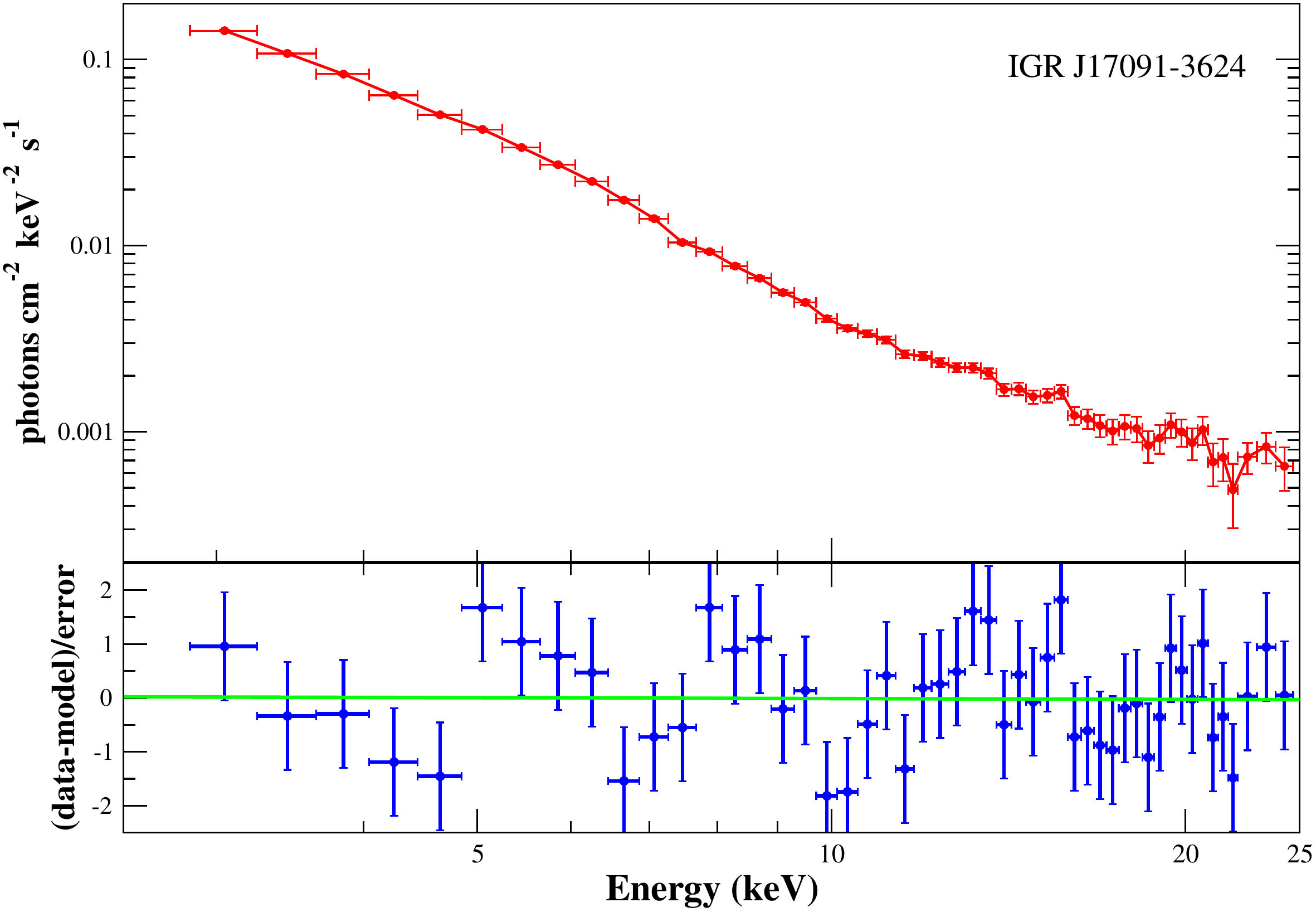}
         \caption{}
     \end{subfigure}
     \hfill
\caption{(a) The 2.5 - 25.0 keV energy spectra fitted with \textsc{tcaf+cutoffpl+pexrav} model in the case of the $\nu$ class data of GRS 1915+105. The residuals are shown in the bottom panel. (b) Energy spectra of the same energy range corresponding to the $\nu$ class data of IGR J17091-3624 fitted with \textsc{tcaf+pexrav} model. Bottom panel contains the residuals.}
\end{figure}

\begin{figure}[h!]
    \centering
    \includegraphics[scale=0.5]{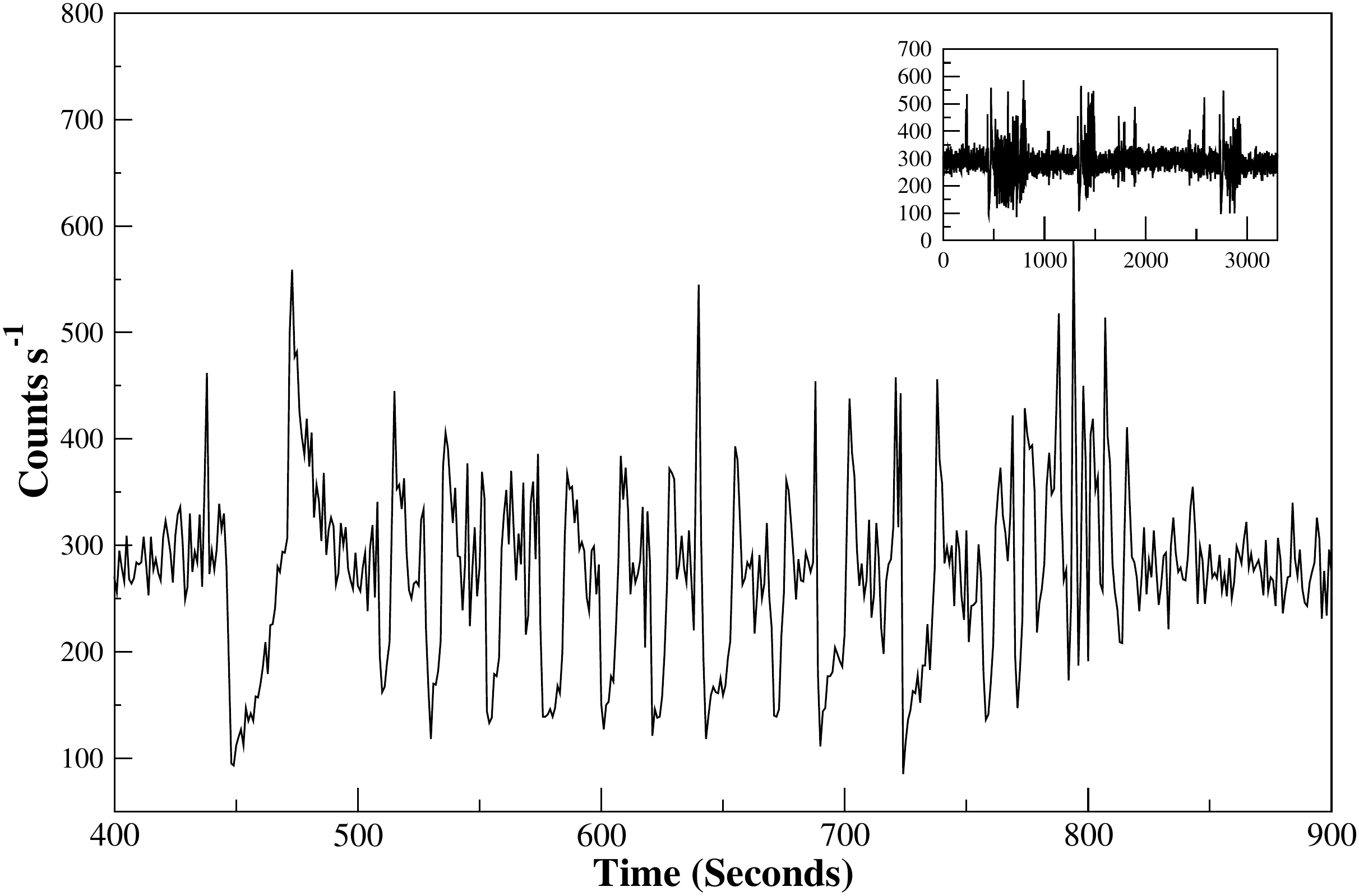}
    \caption{The C2 class lightcurve pertaining to IGR J17091-3624. The source transitions from highly variable phase spanning over few hundred seconds to nearly non-variable phase. The inset contains the full extent of the time series which shows a few variable and non-variable intervals together.}
\end{figure}

\begin{figure}[h!]
    \centering
    \includegraphics[scale=0.5]{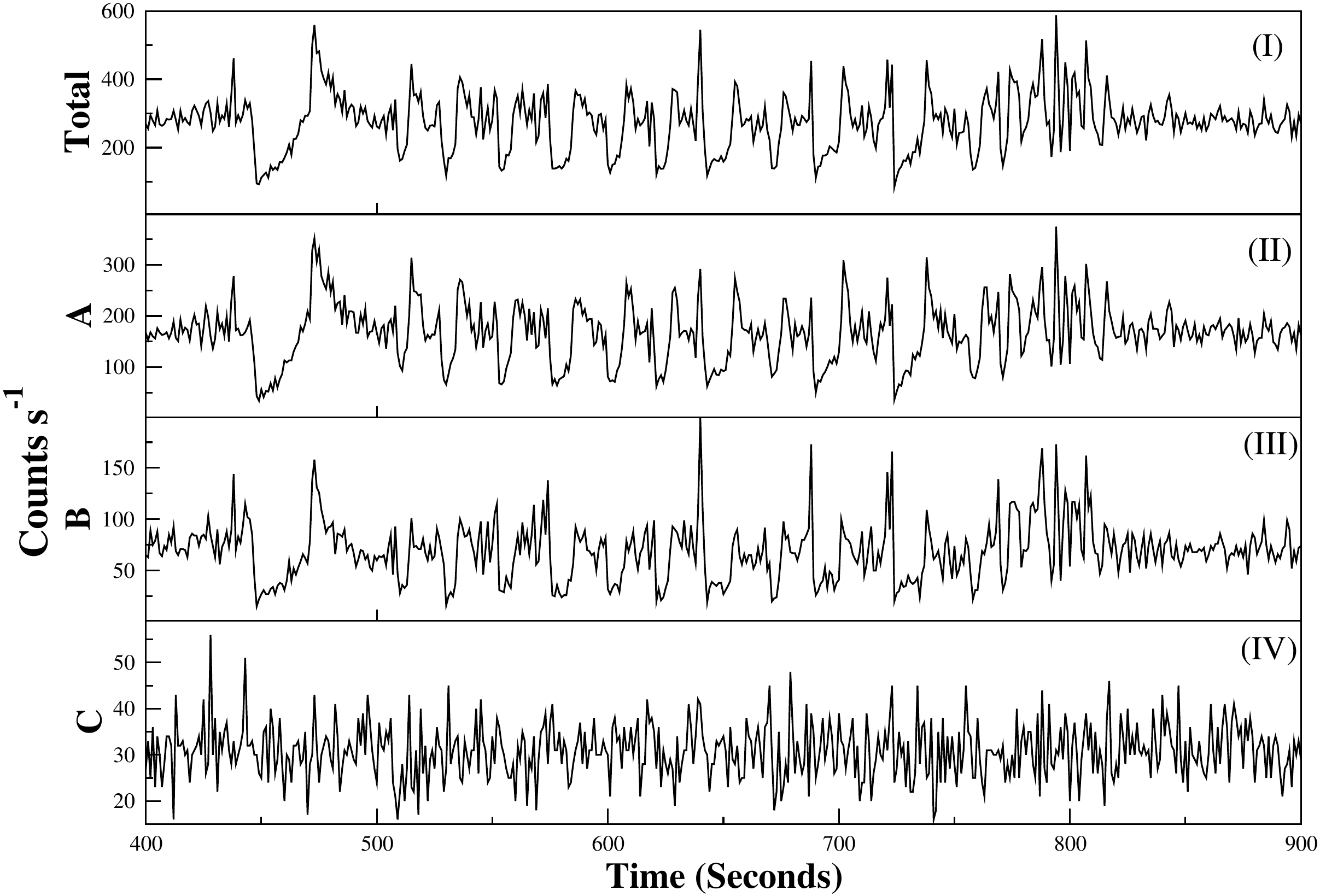}
    \caption{The 2.0-60.0 keV lightcurve of C2 class of IGR J17091-3624 (top panel) as well as the lightcurves corresponding to A (2.0-6.0 keV), B (6.0-15.0 keV) and C (15.0-60.0 keV) bands (panels b-d). The C-band lightcurve remains nearly steady across dip and flare events. }
\end{figure}

\begin{figure}[h!]
    \centering
    \includegraphics[scale=0.45]{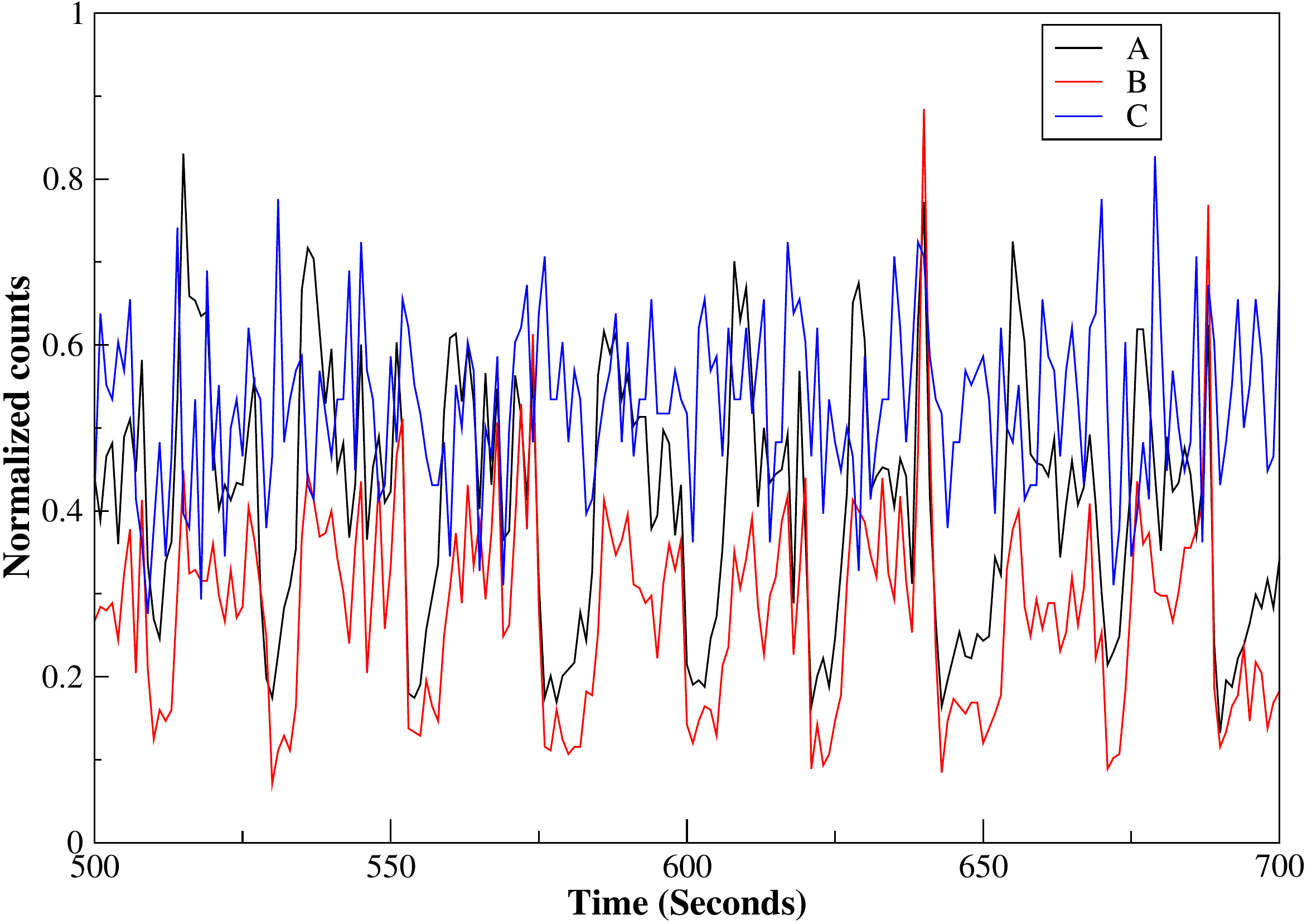}
    \caption{The time variation of the normalized photon count for C2 class of IGR J17091-3624 in A, B and C bands. The B band intensity preferentially decreases during the dips, while the C band intensity remains nearly constant throughout the time interval. }
\end{figure}

\begin{figure}
     \centering
     \begin{subfigure}[b]{0.48\textwidth}
         \centering
         \includegraphics[width=\textwidth]{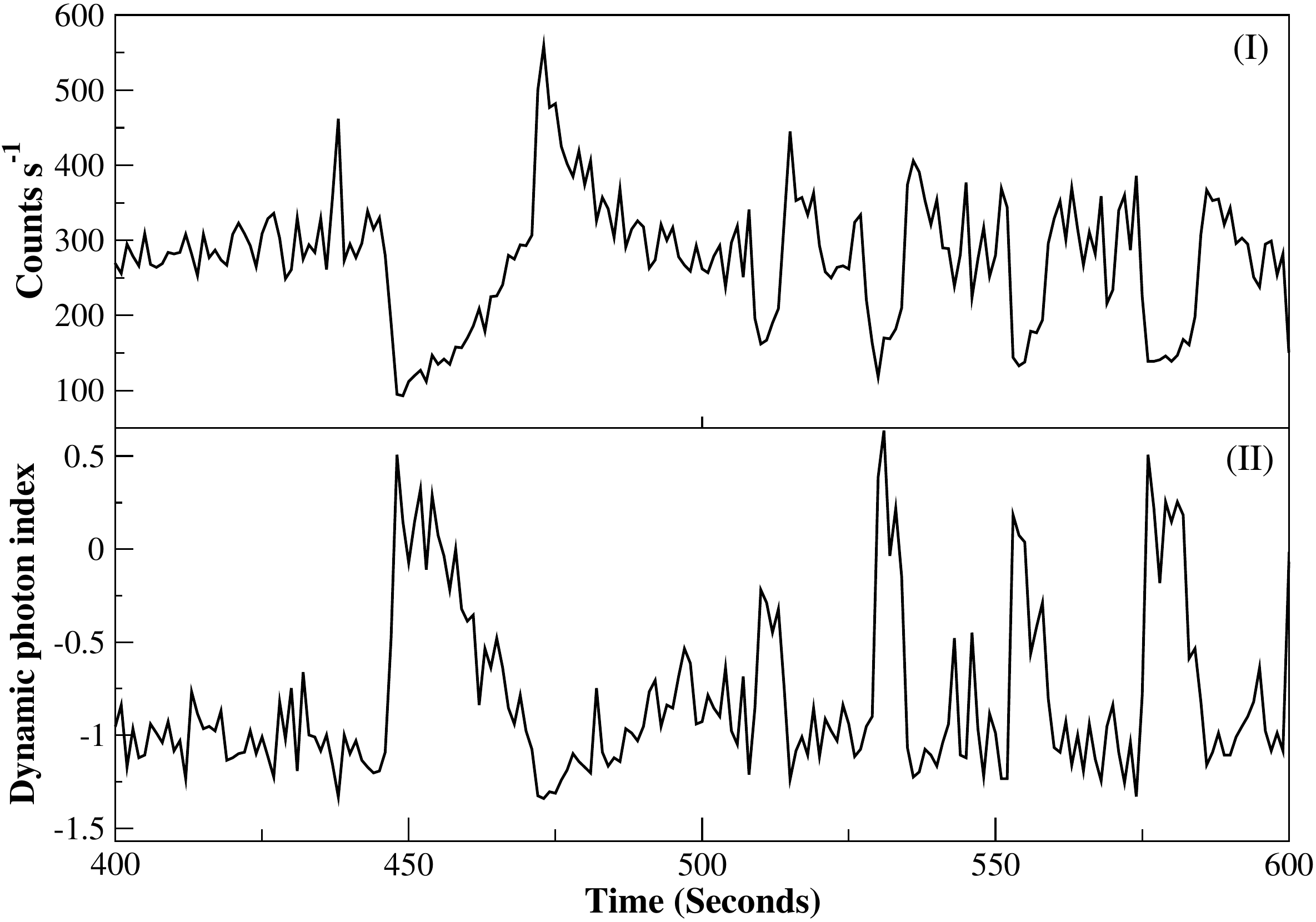}
         \caption{}
     \end{subfigure}
     \hfill
     \begin{subfigure}[b]{0.48\textwidth}
         \centering
         \includegraphics[width=\textwidth]{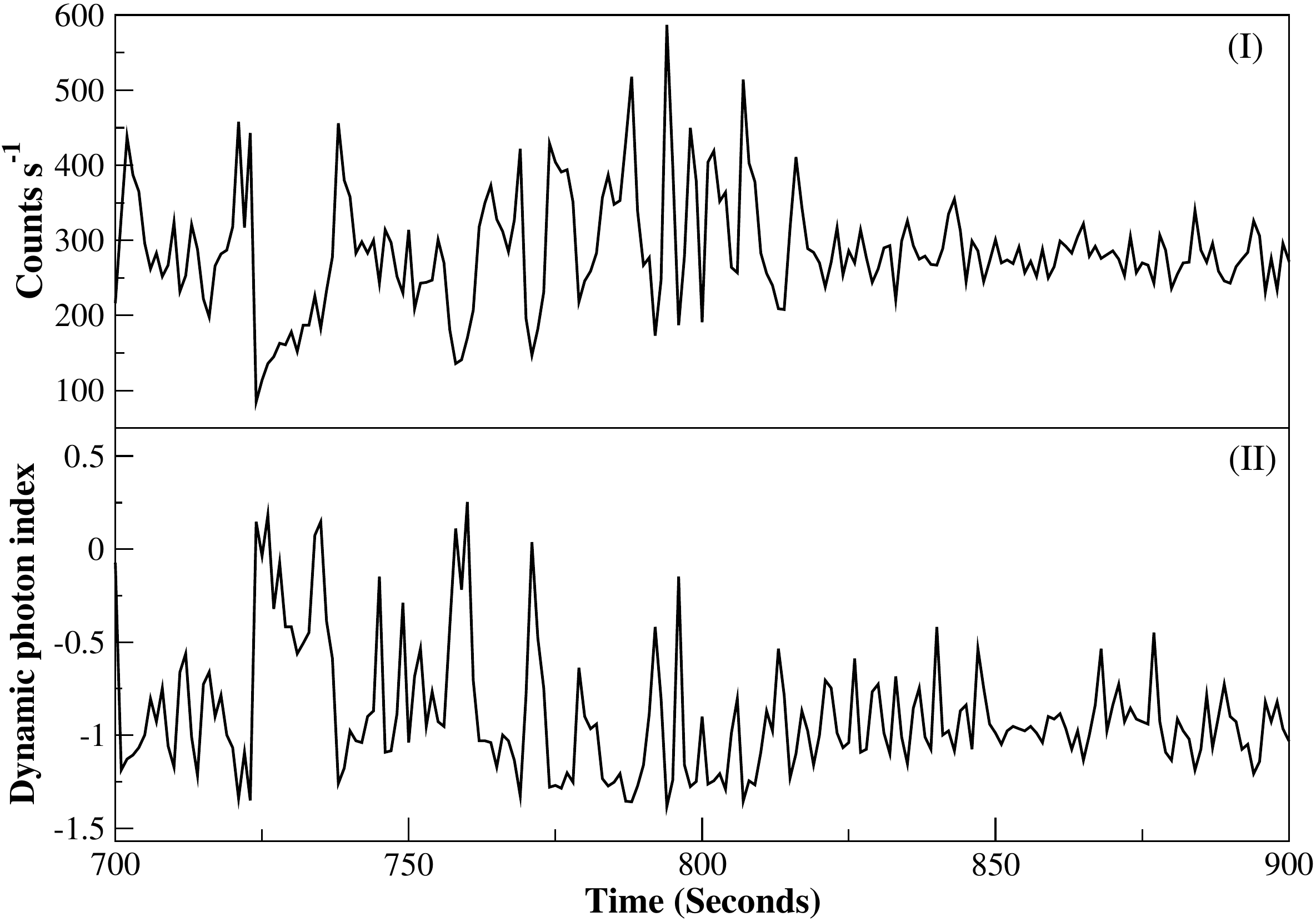}
         \caption{}
     \end{subfigure}
     \hfill
\caption{(a) Time variation of 2.0-60.0 keV photon intensity together with dynamic photon index ($\Theta$) during the variable sub-state of C2 class of IGR J17091-3624. During the dips, $\Theta$ becomes $> 0$, indicating the short transition to harder state. (b) Same set of plots for the transition region from variable to non-variable sub-state. The $\Theta$ value changes gradually during the transition.}
\end{figure}

\begin{figure}[h!]
    \centering
    \includegraphics[scale=0.5]{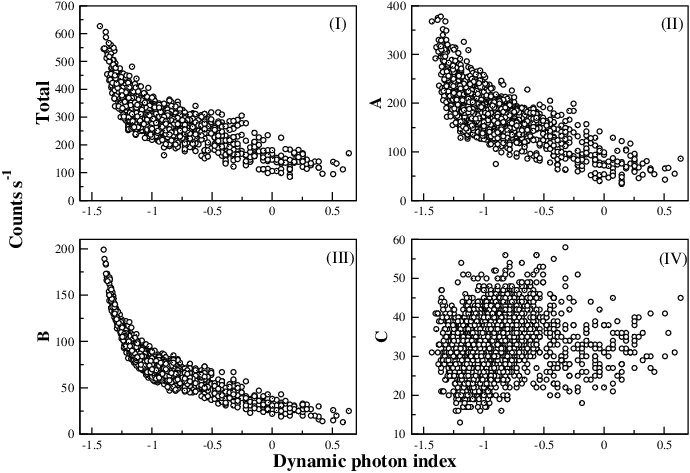}
    \caption{Variation of the total 2.0-60.0 keV photon intensity as well as the A (2.0-6.0 keV), B (6.0-15.0 keV) and C (15.0-60.0 keV) band intensities as a function of dynamic photon index ($\Theta$) in case of C2 class of IGR J17091-3624. Weak positive correlation between photon count and $\Theta$ is observed in C band.}
\end{figure}

\begin{figure}
     \centering
     \begin{subfigure}[b]{0.48\textwidth}
         \centering
         \includegraphics[width=0.85\textwidth,angle=-90]{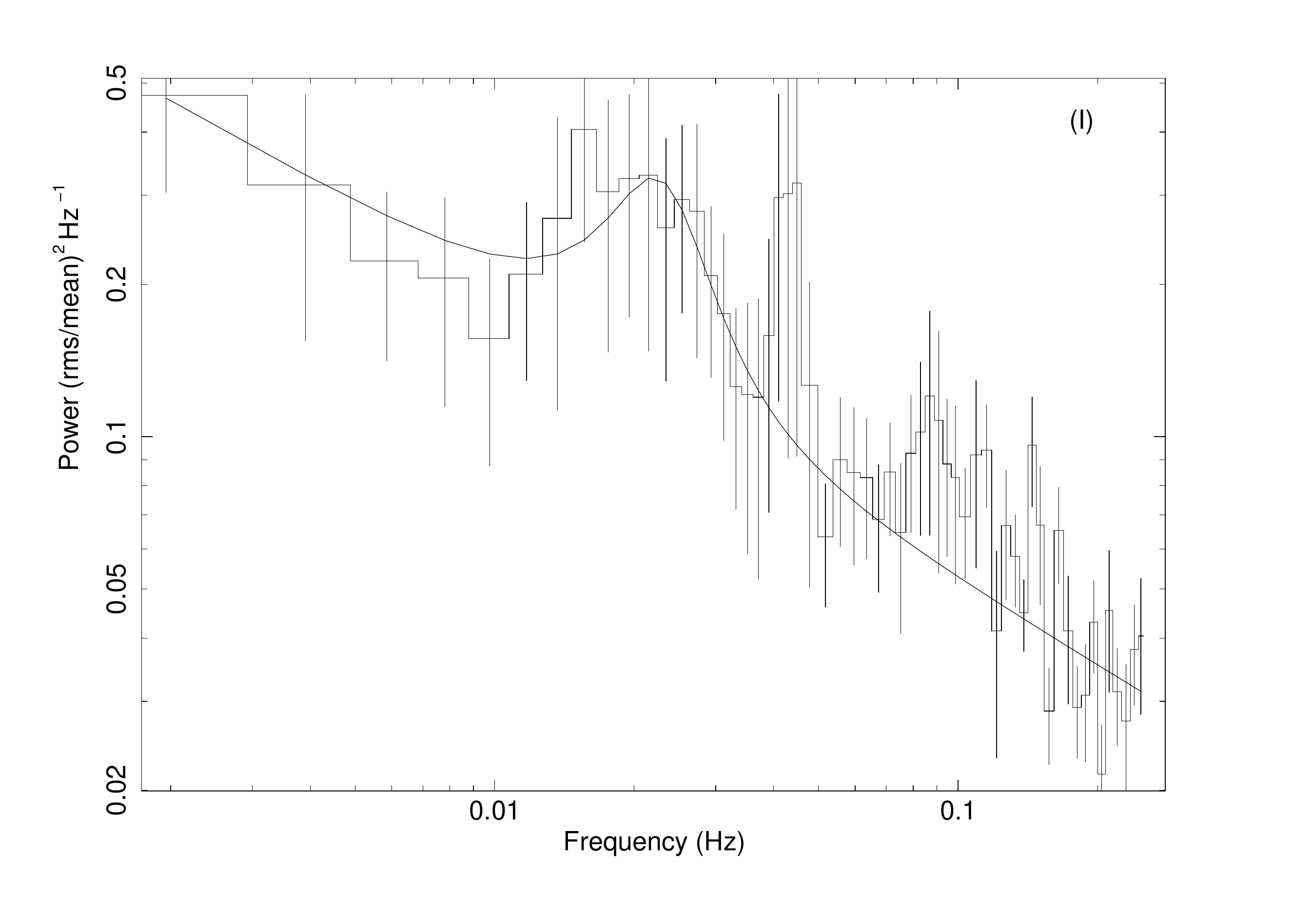}
         \caption{}
     \end{subfigure}
     \hfill
     \begin{subfigure}[b]{0.48\textwidth}
         \centering
         \includegraphics[width=0.85\textwidth,angle=-90]{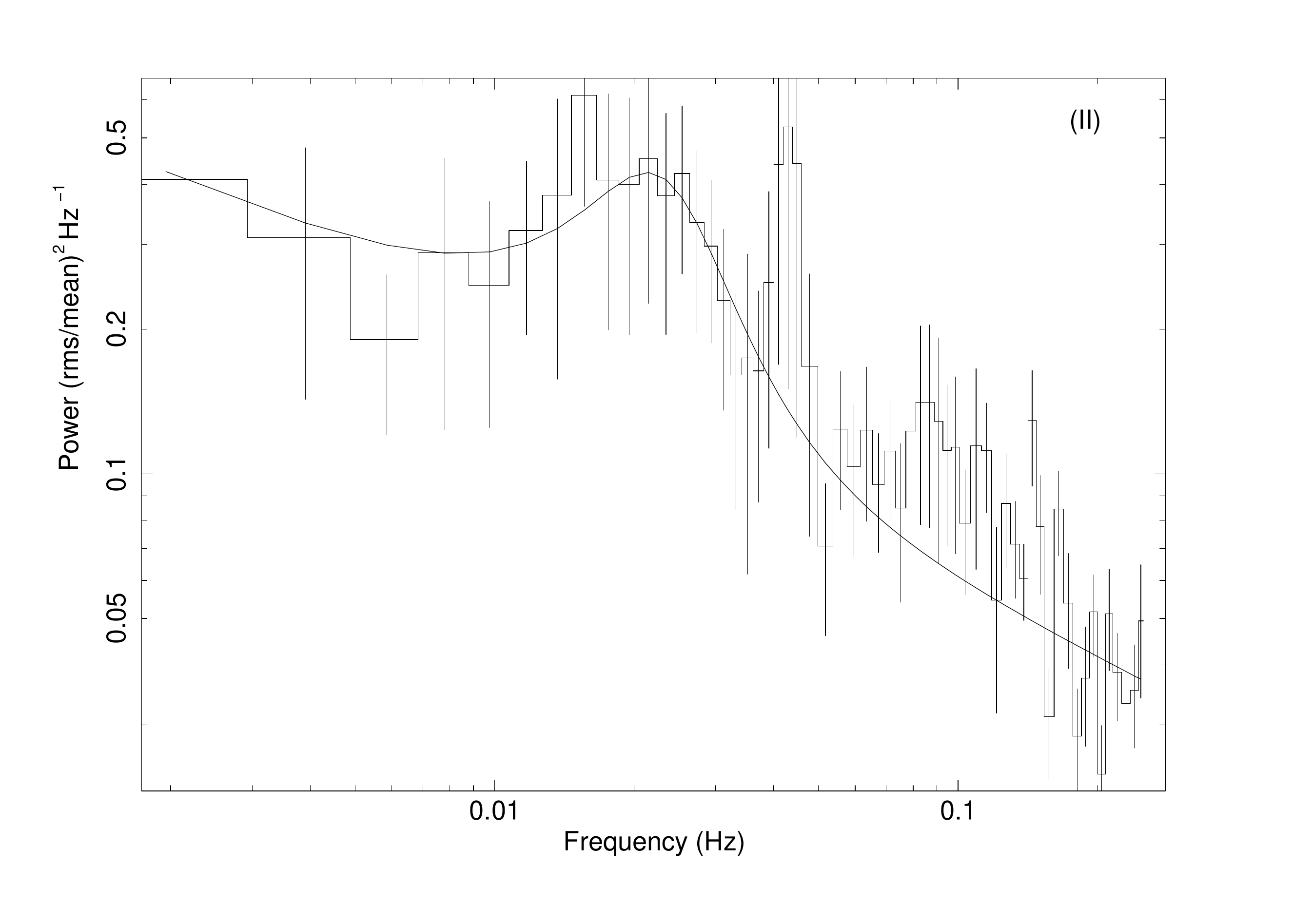}
         \caption{}
     \end{subfigure}
     \hfill
     \begin{subfigure}[b]{0.48\textwidth}
         \centering
         \includegraphics[width=0.85\textwidth,angle=-90]{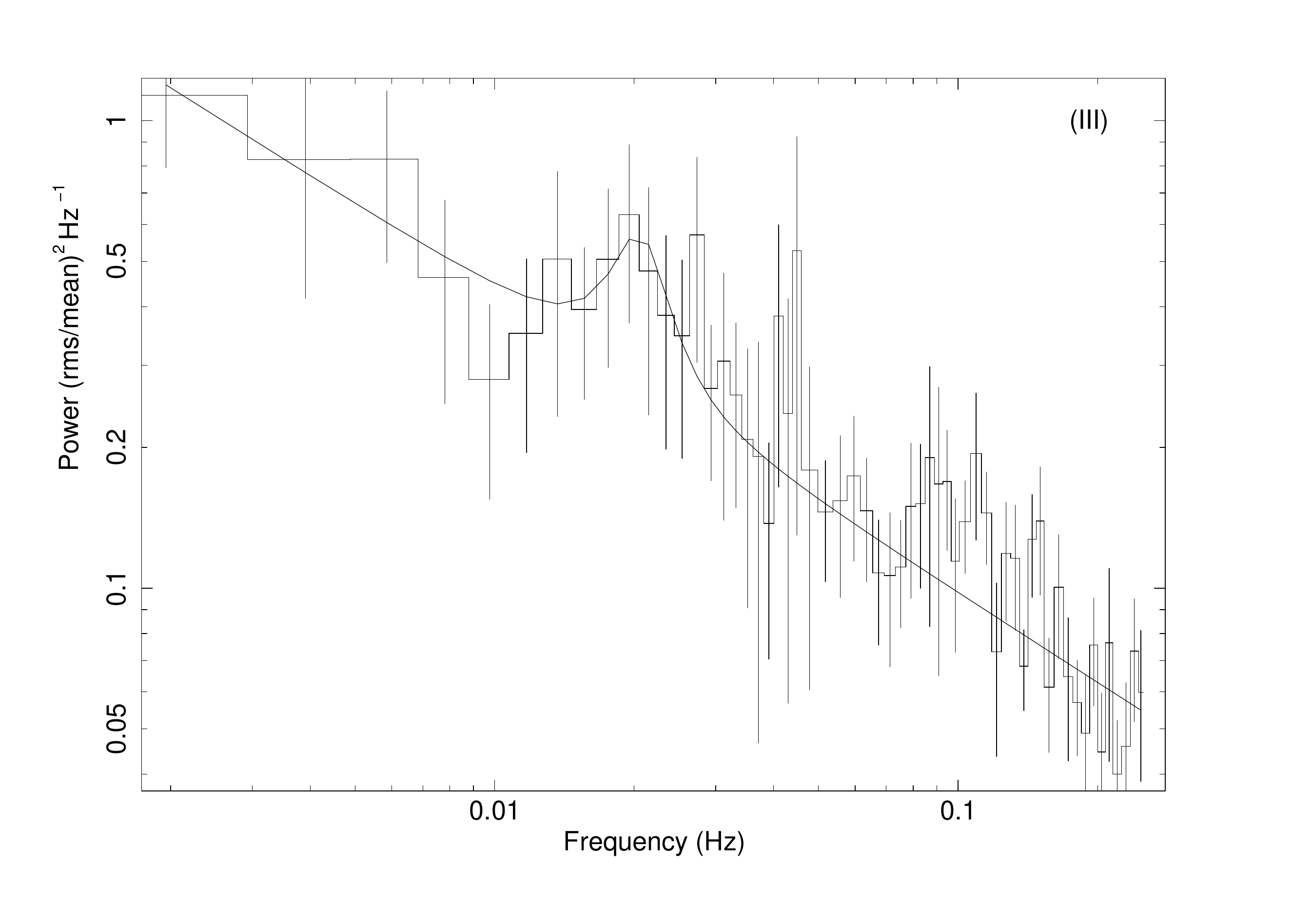}
         \caption{}
     \end{subfigure}
     \hfill
     \begin{subfigure}[b]{0.48\textwidth}
         \centering
         \includegraphics[width=0.85\textwidth,angle=-90]{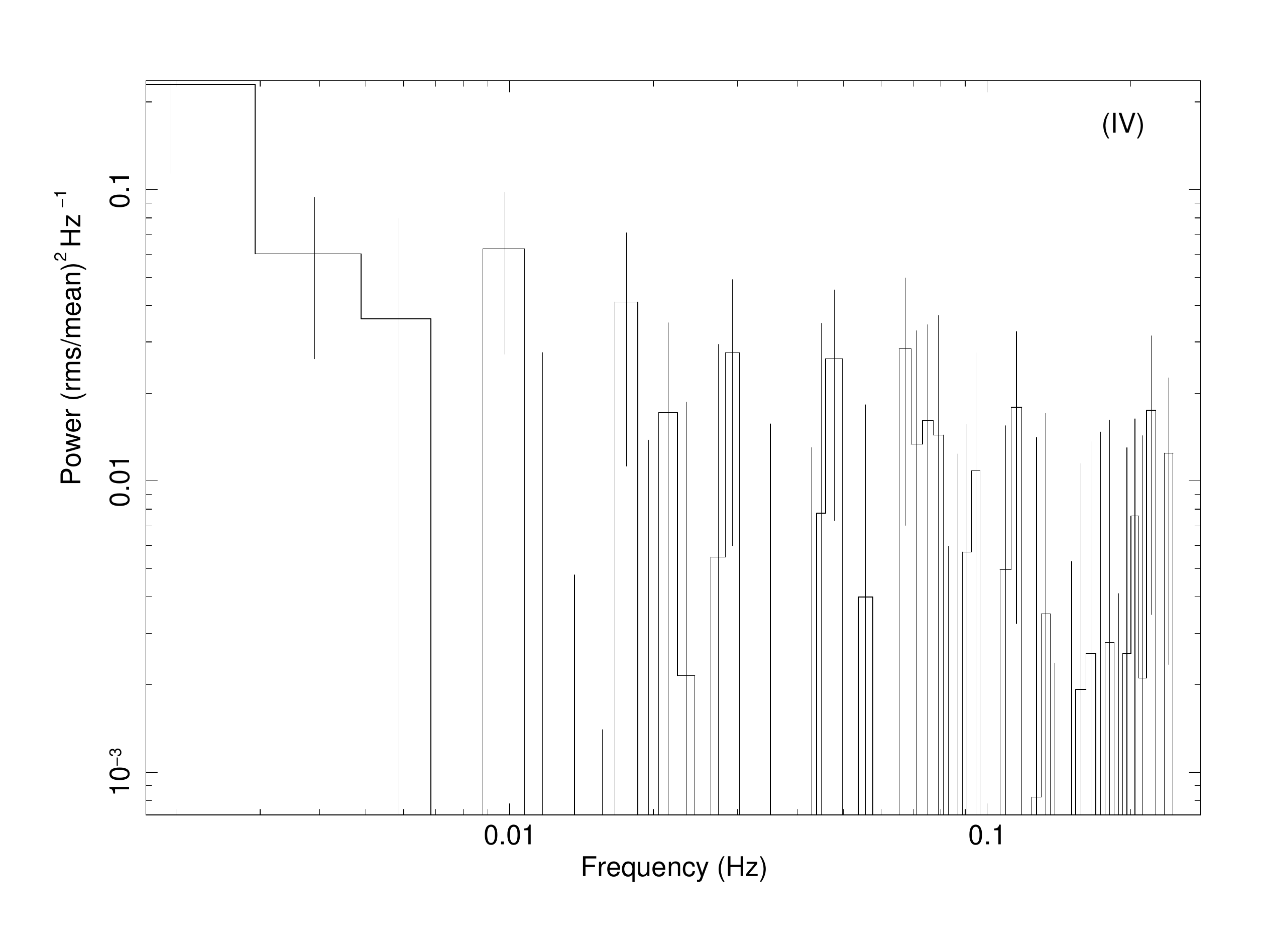}
         \caption{}
     \end{subfigure}
     \hfill
\caption{The fundamental QPO in C2 class of IGR J17091-3624 fitted with Lorentzian profile. In panel (a), the 2.0-60.0 keV PDS is fitted. In panels (b-d), the A, B and C band PDS are fitted. In case of A and B band, broad QPO around 22 mHz is observed. No such peaked component is detected in the case of C-band.}
\end{figure}

\begin{figure}
     \centering
     \begin{subfigure}[b]{0.48\textwidth}
         \centering
         \includegraphics[width=\textwidth]{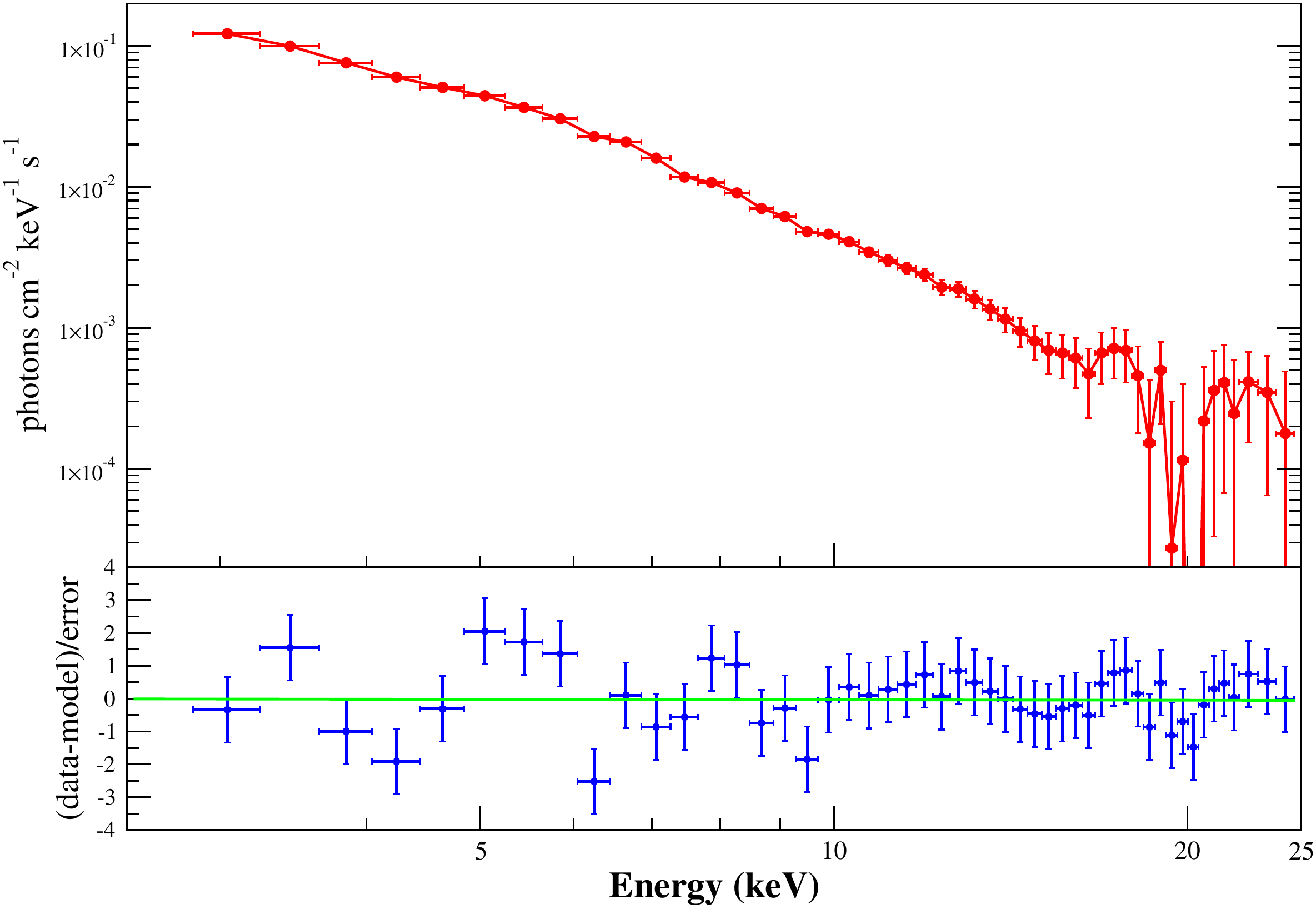}
         \caption{}
     \end{subfigure}
     \hfill
     \begin{subfigure}[b]{0.48\textwidth}
         \centering
         \includegraphics[width=\textwidth]{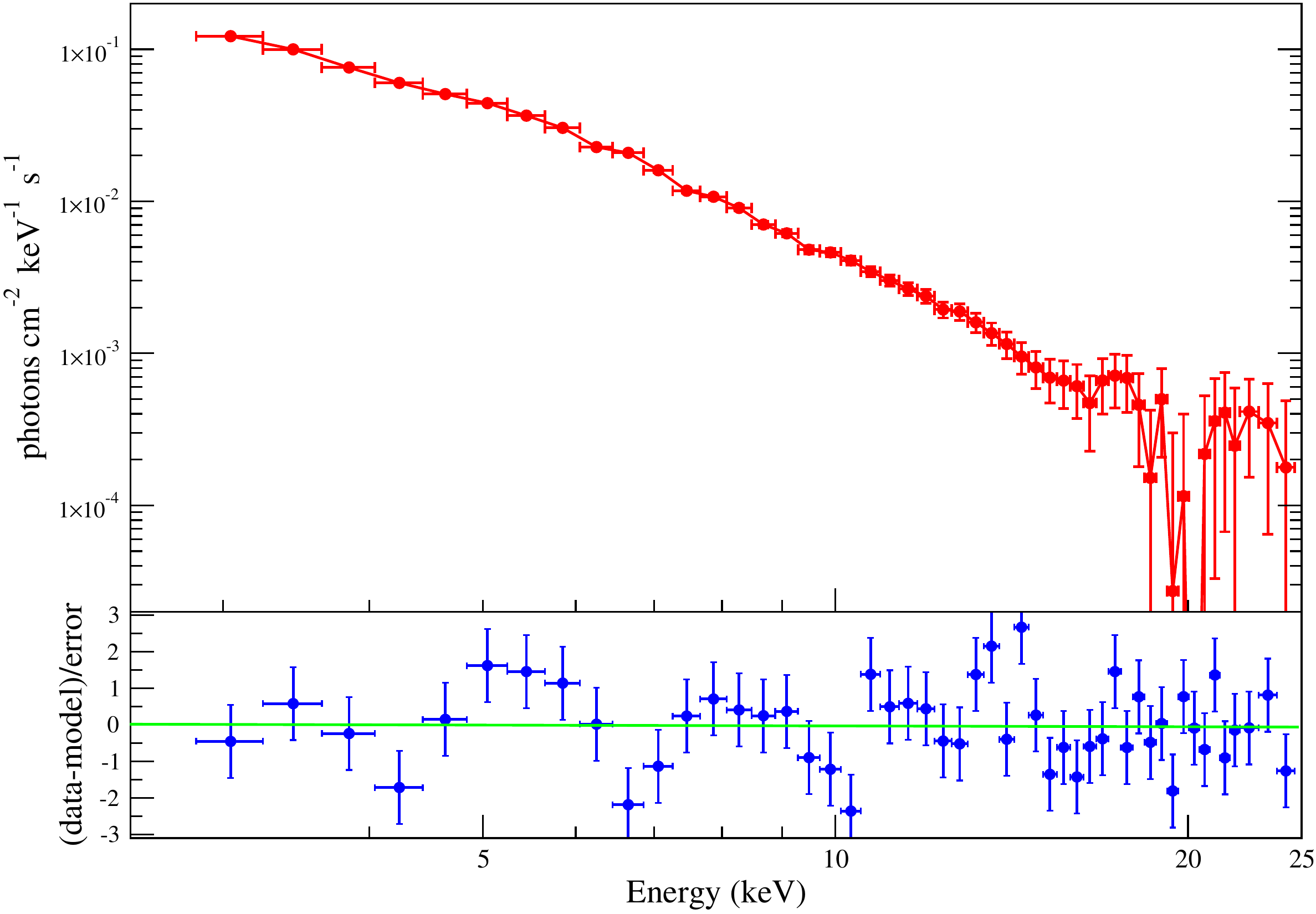}
         \caption{}
     \end{subfigure}
     \hfill
\caption{(a) The 2.5 - 25.0 keV energy spectra fitted with \textsc{tcaf+pexrav} model corresponding to the non-variable phase of C2 class of IGR 17091-3624. The residuals are shown in the bottom panel. (b) Energy spectra for the same energy range corresponding to the highly variable phase of C2 class fitted with \textsc{tcaf+pexrav} model. Bottom panel contains the residuals.}
\end{figure}

\clearpage

\begin{center}

{{\bf Table 3:} Best-fit parameters obtained from the spectral fits corresponding to the steady and variable phase in the C2 class of IGR J17091-3624.}

    \begin{table}[h!]
  \begin{tabular}{l l l l l}
    \hline
    \vspace{0.15cm}
    \multirow{2}{*}{Model parameters} &
     
      \multicolumn{2}{c}{steady phase} &
      \multicolumn{1}{c}{variable phase} \\
    & \textsc{tcaf+pexrav} & \textsc{diskbb+po} & \textsc{tcaf+pexrav} & \textsc{diskbb+po}   \\
    &  & \textsc{+gaussian} &  & \textsc{+gaussian} \\
        \hline
    \vspace{0.2cm}
    
     $T_{\text{in}}$ & -- & $1.65_{-0.07}^{+0.07}$ & -- & $1.51_{-0.06}^{+0.05}$\\
         \vspace{0.2cm}
$\Gamma_{\text{po}}$ & -- & $2.86_{-0.06}^{+0.07}$ & -- & $2.73_{-0.06}^{+0.05}$\\
\hline

         \vspace{0.2cm}
   $\dot{m}_d$ & $0.57_{-0.06}^{+0.05}$ & -- & $0.54_{-0.01}^{+0.02}$ & --   \\
    \vspace{0.2cm}
    $\dot{m}_h$ & $0.25_{-0.05}^{+0.04}$ & -- & $0.23_{-0.02}^{+0.01}$  & -- \\
        \vspace{0.2cm}

    $X_s$ & $15.37_{-1.15}^{+1.10}$ & -- & $27.15_{-1.60}^{+1.55}$  & --  \\
        \vspace{0.2cm}

    $R$ & $1.39_{-0.01}^{+0.02}$ & -- & $2.87_{-0.04}^{+0.05}$  & --\\
\hline
    \vspace{0.2cm}
   $\Gamma_\text{refl}$ & $1.23_{-0.02}^{+0.02}$ & -- & $1.01_{-0.05}^{+0.04}$  & -- \\
 \vspace{0.2cm}
 
    $E_c$ & $2.64_{-0.06}^{+0.07}$ & -- & $2.28_{-0.03}^{+0.03}$   & --\\
    \vspace{0.2cm}
    
    $\text{rel}_\text{refl}$ & $6.36_{-0.07}^{+0.08}$ & -- & $5.08_{-0.08}^{+0.06}$   & --\\
    \hline
    $\chi^2_\text{DOF}$ & $40.99/41$ & $41.30/43$ & $58.91/41$   & $53.37/43$\\
    \hline
  \end{tabular}
  \caption*{Note: $T_{\text{in}}$: inner disk temperature in keV. $\Gamma_{\text{po}}$: photon index from \textsc{diskbb} model. $\dot{m}_d$: disk rate in Eddington unit. $\dot{m}_h$: halo rate in Eddington unit. $X_s$: Shock location in $r_g$. $R$: Shock strength. $\Gamma_\text{refl}$: photon index from \textsc{pexrav}. $E_c$: cutoff energy from \textsc{pexrav}. $\text{rel}_\text{refl}$: reflection fraction.}
\end{table}
\end{center}

\end{document}